\documentclass[traditabstract]{aa} 

\usepackage{amsmath}
\usepackage{amssymb}
\usepackage{enumerate}
\usepackage{colortbl}
\usepackage{graphicx}
\usepackage{txfonts}
\usepackage{natbib}
\usepackage[usenames,dvipsnames]{xcolor}

\bibpunct{(}{)}{;}{a}{}{,} 

\begin{document}

   \title{Distribution of star formation rates during the rapid assembly of NGC~1399 as deduced from its globular cluster system}
   \titlerunning{Distribution of SFRs during the rapid assembly of NGC~1399 as deduced from its GC system}

   \author{C. Schulz \inst{1} \and M. Hilker \inst{1} \and P. Kroupa \inst{2,3} \and J. Pflamm-Altenburg \inst{2} }


   \institute{  \inst{1} European Southern Observatory (ESO), Karl-Schwarzschild-Stra{\ss}e 2, D-85748 Garching bei M\"unchen, Germany \newline
                \email{[cschulz;mhilker]@eso.org} \newline
                \inst{2} Helmholtz-Institut f\"ur Strahlen- und Kernphysik (HISKP), Universit\"at Bonn, Nussallee 14 - 16, D-53115 Bonn, Germany \newline
                \email{[pavel;jpflamm]@astro.uni-bonn.de} \newline
                \inst{3} Charles University in Prague, Faculty of Mathematics and Physics, Astronomical Institute, V Hole\v{s}ovi\v{c}k\'ach 2, CZ-180 00 Praha 8, Czech Republic
             }

 
\abstract{Ultra-compact dwarf galaxies (UCDs) share many properties with globular clusters (GCs) and are found in similar environments. Here, a large sample of UCDs and GCs around NGC~1399, the central giant elliptical of the Fornax galaxy cluster, is used to infer their formation history and also to shed light on the formation of NGC~1399 itself. 

We assumed that all GCs and UCDs in our sample are the result of star cluster (SC) formation processes and used them as tracers of past star formation activities. After correcting our GC/UCD sample for mass loss, we interpreted their overall mass function to be a superposition of SC populations that formed coevally during different formation epochs. The SC masses of each population were distributed according to the embedded cluster mass function (ECMF), a pure power law with the slope $-\beta$. Each ECMF was characterized by a stellar upper mass limit, $M_{\mathrm{max}}$, which depended on the star formation rate (SFR). We decomposed the observed GC/UCD mass function into individual SC populations and converted $M_{\mathrm{max}}$ of each SC population to an SFR. The overall distribution of SFRs reveals under which conditions the GC/UCD sample around NGC~1399 formed. 

Considering the constraints set by the age of the GCs/UCDs and the present stellar mass of NGC~1399, we found that the formation of the GCs/UCDs can be well explained within our framework with values for $\beta$ below 2.3. This finding agrees very well with the observation in young SCs where $\beta \approx 2.0$ is usually found. Even though we took into account that some of the most massive objects might not be genuine SCs and applied different corrections for the mass loss, we found that these considerations do not influence much the outcome. We derived the peak SFRs to be between approximately 300 and 3000 $M_{\odot}\mathrm{yr}^{-1}$, which matches the SFRs observed in massive high-redshift sub-millimeter galaxies and an SFR estimate inferred from NGC~1399 based on the so-called downsizing picture, meaning that more massive galaxies must have formed over shorter periods of time. Our findings give rise to the interpretation that NGC~1399 and its GC/UCD system formed in a relatively short, intense starburst early on.}

\maketitle

\section{Introduction} \label{intro}

The Fornax cluster around the giant elliptical \object{NGC~1399} possesses a very rich system of globular clusters (GCs). \citet{gregg09} estimated $11\,100 \pm 2\,400$ GCs in total within a radius of 0$^{\circ} \!$.9 around \object{NGC~1399}, corresponding to a radius of 320~kpc at a distance of 19~Mpc based on a distance modulus of $(m-M) = 31.4 \pm 0.2$ (\citealt{dirsch03, dirsch04}, see also \citealt{ferrarese00, blakeslee09}). Within $15'$, which corresponds to roughly 83~kpc, \citet{dirsch03} estimated a smaller number of $6\,450 \pm 700$ GCs. Of particular interest is the upper end of the GC mass function, which is dominated by ultra-compact dwarf galaxies (UCDs). The term UCD \citep{phillipps01} is somewhat misleading since these objects are not necessarily dwarf galaxies: They show similarities with nuclei of dwarf galaxies, but they also share many properties with GCs, for which reason two main formation scenarios for UCDs in the Fornax cluster are discussed in the recent literature \citep[e.g.,][]{evstigneeva08, chilingarian08, chilingarian11, mieske12, francis12}:

\begin{enumerate}[(a)]
\item UCDs are dynamically evolved nucleated dwarf galaxies, from which outer stellar components were removed while orbiting in the center of the Fornax galaxy cluster and suffering from its strong tidal gravitational field \citep[threshing scenario; e.g.,][]{bekki01, bekki03, drinkwater03, goerdt08, thomas08, pfeffer13, pfeffer14, pfeffer16}.
\item UCDs are the brightest GCs of the rich \object{NGC~1399} globular cluster system and thus they are the result of star cluster formation processes \citep[e.g.,][]{mieske02, mieske12}. Moreover, it has been proposed that the very massive UCDs could also form in the so-called merged star cluster scenario, where massive complexes of star clusters merge and thereby form a super star cluster \citep{kroupa98, fellhauer&kroupa02, mieske06, bruens11, bruens12}.
\end{enumerate}

Although almost two decades of research have passed since the discovery of UCDs \citep{minniti98, hilker99, drinkwater00}, their nature is still puzzling \citep[e.g.,][]{phillipps13}: Various investigations found evidence for both formation scenarios, but could not confirm either of the hypotheses with certainty \citep[e.g.,][]{mieske04, evstigneeva08, wittmann16}. However, there is growing evidence that UCDs are rather a "mixed bag of objects" \citep{hilker09book} than a distinct type of object \citep[e.g.,][]{chilingarian11}. 

In the Fornax galaxy cluster, GCs have effective radii from smaller than 1 and up to 10~pc with an average of around 3~pc \citep{masters10, puzia14}. Their masses range from $10^4~M_{\odot}$ up to lower than $10^7~M_{\odot}$ \citep[][their Fig.~15]{puzia14}. UCDs, on the other hand, have some overlap with GCs, but also extend the parameter space to larger sizes and masses: their effective radii range from a few pc to up to 100~pc \citep[e.g.,][]{drinkwater03, evstigneeva07, evstigneeva08, hilker07, mieske08}, while their masses lie between $10^6~M_{\odot}$ and lower than $10^8~M_{\odot}$ \citep[e.g.,][]{drinkwater03, evstigneeva07, hilker07, mieske08, frank11}, bridging the region between classical GCs and compact elliptical galaxies.

In the literature, several arguments support that most of the Fornax UCDs are very massive GCs:

\begin{itemize}
\item The luminosities of GCs and UCDs are distributed smoothly and their luminosity functions do not show any bimodality \citep{mieske02, mieske04}. Furthermore, the UCDs in Fornax are consistent with being drawn from the bright tail of the GC luminosity function. However, a small excess at the bright end is statistically possible \citep{mieske04, gregg09, mieske12}.
\item GCs and UCDs form a continuum in the luminosity-size plane \citep{mieske06}. 
\item UCDs exhibit the full range of (high and low) metallicities as observed for GCs \citep{francis12}.
\item The spread of age and metallicity of the UCDs is consistent with that observed for GCs \citep{francis12}.
\item Most of the UCDs have super-solar $\alpha$-element abundances, implying short formation times, similar to those of GCs \citep{francis12}.
\end{itemize}

However, even if the UCDs were not genuine GCs, several findings suggest that they are at least the result of a star cluster (SC) formation process:

\begin{itemize}
\item \citet{kissler-patig06} placed young massive clusters (YMCs) with masses higher than $10^7 M_{\odot}$ on three different scaling relations and found their positions to be similar to those of the UCDs, in particular for the most massive YMCs.
\item UCDs have metallicities close to but slightly below those of YMCs of comparable masses \citep{mieske06}.
\item Fitting a simple stellar population model to the spectra of UCDs reveals that UCDs are in agreement with a pure stellar content \citep{hilker07, chilingarian11} so that no DM component is needed in UCDs within their 1--3 half-mass radii \citep{hilker07}. \citet{chilingarian11} found almost all UCDs to be compatible with no DM inside. More recently, \citet[their Table~3]{mieske13} found that only the most massive UCDs require an additional mass component to compensate the elevated $M/L$ ratio, which they suggested might be massive black holes, while the lower-mass UCDs do not need any form of additional, non-luminous matter \citep[see also][]{dabringhausen09, dabringhausen10, dabringhausen12}.
\end{itemize}

Following \citet{mieske12}, who found that most UCDs are compatible with being formed in the same way as GCs, we assume that it is justified to treat UCDs, like GCs, as (very) massive SCs and assume that they formed in SC formation processes. We aim to determine the conditions under which this occurred and what it indicates about the assembly history of NGC~1399.

However, as mentioned before, even if the majority of the UCDs are compatible with being giant ancient SCs, it is very likely that some of the most massive UCDs did not form in an SC formation process. For instance, some of the UCDs exhibit extended surface brightness profiles or even tidal features \citep{richtler05, voggel16} or appear asymmetric or elongated \citep{wittmann16}. In addition, \citet{voggel16} detected for the first time the tendency of GCs to cluster around UCDs. In the Virgo galaxy cluster, a fraction of the very massive UCDs are thought to be of galaxy origin \citep[e.g.,][]{strader13, seth14, liu15a, liu15b, norris15}, while a fraction of the faintest (and thus lowest mass) UCDs might instead be related to compact SCs \citep{brodie11}. Thus, in two additional scenarios we examine how the distribution of required SFRs changes if some of the most massive UCDs in Fornax are excluded from the GC/UCD mass distribution.

\section{Aim and structure of this paper} \label{sect_aim}

We introduce the underlying theory in Sect.~\ref{sect_framework} and define there the so-called embedded cluster mass function (ECMF). The ECMF describes the stellar mass function of a population of SCs at their birth and is characterized by a power-law behavior up to an individual stellar upper mass limit, $M_{\mathrm{max}}$. The latter is determined by the star formation rate (SFR) according to the SFR-$M_{\mathrm{max}}$ relation \citep{weidner04, randria13}. In Sect.~\ref{sect_sample} we present the available data on the mass distribution of GCs/UCDs in the central Fornax galaxy cluster and construct an overall GC/UCD mass function from it.

The ECMF and the observed GC/UCD mass function are related to each other in the following way: We assume that every object of the GC/UCD sample formed in an SC formation process and is thus referred to as an SC. Each SC forms together with many other SCs of similar age in a population that is described by the ECMF. By this, we implicitly assume that each of those GC/UCD populations form in the same way as it is observed in the local Universe today \citep[e.g.,][]{fall_rees77, okazaki95, fall_zhang01, elmegreen10}. Accumulating all GC/UCD populations ever formed in different formation epochs is equivalent to a superposition of all corresponding ECMFs, which leads to their overall birth mass function. SCs are subject to changes, particularly in mass, as the SCs interact with the environment and the stars in these SCs become older. Both lead to mass loss in the course of time, which transforms the natal GC/UCD mass function to the present-day GC/UCD mass function.

To learn under which conditions the GCs/UCDs in the present-day mass function formed, two investigation steps are required: First, it is necessary to determine how aging affected each GC/UCD. We are particularly interested in quantifying how much mass each of them lost since its birth. We explain in Sect.~\ref{sect_corrections} which corrections need to be applied to the observed GC/UCD sample. This knowledge will enable us to reconstruct the natal GC/UCD mass function from their present-day mass function. Second, the natal GC/UCD mass function can then be decomposed into individual SC populations described by the ECMF. This is carried out in Sect.~\ref{sect_decomposition}. Since the ECMF implicity depends on the SFR through the upper mass limit, $M_{\mathrm{max}}$, we will be able to determine which distribution of SFRs led to the formation of such a rich sample of GCs and UCDs around NGC~1399. The determined SFR distributions are presented in Sect.~\ref{sect_sfr.distr}. We continue with a discussion of our assumptions, the results, and the implications for the formation of NGC~1399 in Sect.~\ref{sect_discussion} and conclude our investigations in Sect.~\ref{sect_concl}.

\section{Underlying framework} \label{sect_framework}

We assume that a large sample of SCs like the distibution of GCs and UCDs around NGC~1399 is the product of many individual formation epochs. The overall GCs/UCDs mass distribution function can be described by a superposition of many different ECMFs, each describing the mass distribution function of one individual GC/UCD population. The goal here is to decompose the overall GCs/UCDs mass function into separate ECMFs to reveal under which conditions each population formed.

This work is based on \citet{schulz15}. All necessary ingredients for this paper are briefly reviewed in this section. For all details, we refer to Schulz et al. (2015).

\subsection{Embedded cluster mass function (ECMF)} \label{sub_ecmf}

The ECMF describes the mass distribution of a population of newly born SCs that were formed coevally from the same parent molecular cloud during one SC formation epoch of length $\delta t$:

   \begin{equation} \label{ecmf}
      \xi_{\mathrm{ECMF}} (M) = \frac{\mathrm{d} N_{\mathrm{ECMF}}}{\mathrm{d} M} = k \left( \frac{M}{M_{\mathrm{max}}} \right)^{- \beta} ,
   \end{equation}

\noindent with the stellar upper mass limit for SCs, $M_{\mathrm{max}}$, the normalization constant $k$, and the index $\beta$ of the ECMF, which lies typically in the range $1.6 \lesssim \beta \lesssim 2.5$. In our framework \citep[see][]{schulz15}, $k$ is determined by

   \begin{equation} \label{norm_k}
      k = (\beta - 1) ~ M_{\mathrm{max}}^{-1} .
   \end{equation}

\noindent With this, the total number of SCs, $N_{\mathrm{ECMF}}$, and the total mass, $M_{\mathrm{ECMF}}$, of one SC population can directly be calculated from the ECMF (Eq.~(\ref{ecmf})):

   \begin{equation} \label{N_ecmf}
      N_{\mathrm{ECMF}} = \int^{M_{\mathrm{max}}}_{M_{\mathrm{min}}} { \xi_{\mathrm{ECMF}} (M) ~ \mathrm{d} M } = \left( \frac{M_{\mathrm{max}}}{M_{\mathrm{min}}} \right)^{\beta - 1} - 1 ,
   \end{equation}

   \begin{align} \label{M_ecmf} 
\begin{split}
      M_{\mathrm{ECMF}} & = \int^{M_{\mathrm{max}}}_{M_{\mathrm{min}}} { M ~ \xi_{\mathrm{ECMF}} (M) ~ \mathrm{d} M } \\ & =
  \begin{cases}
 M_{\mathrm{max}} ~ \left( \ln M_{\mathrm{max}} - \ln M_{\mathrm{min}} \right)  &, \ \beta = 2 \\
 M_{\mathrm{max}} ~ \left[ \frac{\beta - 1}{2 - \beta} ~ \left( 1 - \left( \frac{ M_{\mathrm{min}} }{ M_{\mathrm{max}} } \right)^{2 - \beta} \right) \right]  &, \ \beta \neq 2 . \\
  \end{cases}
\end{split}
   \end{align}

\noindent $M_{\mathrm{min}}$ is the stellar lower mass limit for an embedded SC or a group of stars. For all following computations, we assume $M_{\mathrm{min}} = 5~M_{\odot}$ \citep[cf.][]{weidner04, schulz15}.

The length of one SC formation epoch, $\delta t$, corresponds to the timescale over which the interstellar medium forms molecular clouds from which a new population of embedded SCs emerges. This time span lies between at least a few Myr and at most a few 10 Myr \citep[e.g.,][]{fukui99, yamaguchi01, tamburro08, egusa04, egusa09} and was also determined in \citet[their Table~2]{schulz15}. During the time $\delta t$, the total mass of the SC population, $M_{\mathrm{ECMF}}$, is formed at a constant SFR,

   \begin{equation} \label{mtotsfrdt}
      M_{\mathrm{ECMF}} = \mathrm{SFR} \cdot \delta t .
   \end{equation}

\noindent Combining Eq.~(\ref{M_ecmf}) and Eq.~(\ref{mtotsfrdt}), we obtain the following relation between the SFR and stellar upper mass limit for SCs $M_{\mathrm{max}}$:

\begin{align} \label{sfr}
 \mathrm{SFR} &= 
  \begin{cases}
 \frac{M_{\mathrm{max}}}{\delta t} ~ \left( \ln M_{\mathrm{max}} - \ln M_{\mathrm{min}} \right)  &, \ \beta = 2 \\
 \frac{ M_{\mathrm{max}} }{\delta t} ~ \frac{\beta - 1}{2 - \beta} ~ \left( 1 - \left( \frac{ M_{\mathrm{min}} }{ M_{\mathrm{max}} } \right)^{2 - \beta} \right)  &, \ \beta \neq 2 . \\
  \end{cases}
\end{align}

\noindent This describes the observational SFR-$M_{\mathrm{max}}$ relation by \citet[see also \citealt{randria13}]{weidner04} very well \citep[][their Fig.~4]{schulz15}. According to this, during high-SFR episodes SCs of higher masses are formed than at low-SFR episodes. This means in turn that high SFRs are essential for the formation of high-mass SCs. For the ECMF, this introduces an implicit dependence on the SFR since its upper limit, $M_{\mathrm{max}}$, is a function of the SFR:

   \begin{equation} \label{ecmf_limits}
      \xi_{\mathrm{ECMF}} (M) \equiv \xi_{\mathrm{ECMF,SFR}} ( M_{\mathrm{min}} \le M \le M_{\mathrm{max}} (\mathrm{SFR}) ) .
   \end{equation}

Usually, the theoretical upper mass limit for SCs, $M_{\mathrm{max}}$, is barely known, while it is easier to determine the mass of the physically most massive SC, $M_{\mathrm{SC,max}}$. In our framework, these two quantities are related as follows:

   \begin{align} \label{M_eclmax}
\begin{split}
      M_{\mathrm{SC,max}} & = 
  \begin{cases}
  ~ \left( \ln 2 \right) ~ M_{\mathrm{max}}  &, \ \beta = 2 \\
  ~ \frac{\beta - 1}{2 - \beta} ~ \left( 1 - 2^{\frac{2 - \beta}{1 - \beta}} \right) ~ M_{\mathrm{max}}  &, \ \beta \neq 2 , \\
  \end{cases}
\end{split}
   \end{align}

\noindent so that Eq.~(\ref{sfr}) can be rewritten as a function of $M_{\mathrm{SC,max}}$.

\subsection{Concept of the integrated galactic embedded cluster mass function (IGECMF)} \label{sub_igecmf}

Star cluster formation typically continues over more than just one formation epoch, $\delta t$. In our framework, this means that for each formation epoch one fully populated ECMF is added to the already existing sample of SCs. An observed sample of SCs is thus (unless the SCs are coeval) a superposition of several SC populations, each described by the ECMF. When the star formation history (SFH) or at least the distribution of SFRs is known, an individual $M_{\mathrm{max}}$ can be determined for each formation epoch based on Eq.~(\ref{sfr}). This fully determines each ECMF (Eq.~(\ref{ecmf})) through Eq.~(\ref{norm_k}). Finally, all ECMFs have to be summed to gain the IGECMF, which is the overall mass distribution function of all SCs ever formed during the considered formation episode. A visual example of this concept is given in \citet[their Fig.~3]{schulz15}. We note that the superposition of power-law ECMFs each generated during one SC formation epoch yields IGECMFs with a Schechter-like turn-down at the high-mass end \citep[their Fig.~9]{schulz15}.

Mathematically, this summing can be expressed by an integral over all possible SFRs,

   \begin{equation} \label{igecmf}
      \xi_{\mathrm{IGECMF}} (M) = \int_{\mathrm{SFR_{min}}}^{\mathrm{SFR_{max}}} { \xi_{\mathrm{ECMF,SFR}} (M) ~ F(\mathrm{SFR}) ~ \mathrm{d} \mathrm{SFR} } ,
   \end{equation}

\noindent where the ECMF is modulated by the distribution function of SFRs, $F(\mathrm{SFR})$. The latter reveals the number of SC formation epochs (SCFEs) $\mathrm{d} N_{\mathrm{SCFE}} (\mathrm{SFR})$ per SFR interval:

   \begin{equation} \label{F}
      F(\mathrm{SFR}) = \frac{\mathrm{d} N_{\mathrm{SCFE}} (\mathrm{SFR}) }{\mathrm{d} \mathrm{SFR}} .
   \end{equation}

\noindent With $F(\mathrm{SFR})$, only those ECMFs are selected to contribute to the IGECMF whose $M_{\mathrm{max}}$ correspond to an SFR that appeared in the SFH.

\subsection{Optimal sampling} \label{sub_sampling}

In the following we treat SCs individually, therefore it is necessary to know their individual masses. This requires the sampling of individual SC masses from the ECMF. It is possible to interpret the ECMF as a probability density distribution function such that SC formation becomes a probabilistic or stochastic process \citep[e.g.,][]{bastian08}. However, observations suggest that the formation of SCs is self-regulated instead of probabilistic or stochastic \citep[][see also references therein]{bate09, pflamm-altenburg13, kroupa15review}: Data of young most massive SCs in galaxies with high SFRs obtained by \citet{randria13} show the dispersion of the $M_{\mathrm{max}}$ values to be significantly smaller at a given galaxy-wide SFR than the probabilistic interpretation would predict. Furthermore, \citet{pflamm-altenburg13} found that the masses of the most massive SCs in the galaxy M33 decline with galactocentric radius following the gas surface density, which is not expected from a purely probabilistic model. This is why we decided to use a deterministic sampling technique, called improved optimal sampling \citep{schulz15}.

The idea is to divide an ECMF into pieces such that each of them corresponds to one individual SC. Mathematically, this requires the two following conditions to be fulfilled:

   \begin{equation} \label{cond1}
      1 = \int^{m_i}_{m_{i+1}} { \xi_{\mathrm{ECMF}} (M) ~ \mathrm{d} M } ,
   \end{equation}

   \begin{equation} \label{condM} 
      M_i = \int^{m_i}_{m_{i+1}} { M ~ \xi_{\mathrm{ECMF}} (M) ~ \mathrm{d} M } .
   \end{equation}

\noindent Equation~(\ref{cond1}) ensures that there is exactly one SC within the integration limits, while Eq.~(\ref{condM}) determines the mass of this SC. We note that the integration limits in these two equations must be the same and within the global limits $M_{\mathrm{min}}$ and $M_{\mathrm{max}}$. The evaluation must start with $i = 1$ at the high-mass end, $m_1 = M_{\mathrm{max}}$, and must be continued toward lower masses. Thus, with increasing number $i$, the SCs become less massive.

\section{Observed GC and UCD samples in Fornax} \label{sect_sample}

Our analysis bears several challenges: the majority of GCs and UCDs in the Fornax galaxy cluster are very old (age estimates range between roughly 8 and 15~Gyr; \citealt{kundu05, hempel07, firth09, chilingarian11, francis12}). Above an age of about 5~Gyr, age measurements become less certain because isochrones lie closer to each other in line index diagrams the higher the probed age. Thus, it is impossible to tell when exactly each individual object formed. However, even if we cannot draw firm conclusions about the formation history of each and every GC/UCD, we can at least make a statistically reliable statement about the formation of the whole GC/UCD system based on the sheer number of objects: \citet{gregg09} estimated 11$\,$100 $\pm$ 2$\,$400 GCs/UCDs within 320~kpc around NGC~1399. However, only a fraction of them have been spectroscopically confirmed so far. As shown by \citet[their Fig.~3]{mieske12}, very many GCs and UCDs are confirmed within a radius of 50~kpc around NGC~1399 thanks to a high spatial coverage. At radii between 50~kpc and 100~kpc, there are fewer, while beyond 100~kpc the number of confirmed GCs/UCDs decreases strongly. This is not only caused by their decreasing radial number density profile \citep[e.g.,][their Fig.~15]{schuberth10} but also due to an incomplete spatial coverage inherent to spectroscopic surveys. Thus, it is not straightforward to obtain a statistically representative sample around NGC~1399. We try to achieve this by combining a spectroscopic and a highly confident photometric sample.

\paragraph{Spectroscopic sample ('spec' sample):}
Our first sample contains 935 of the brightest GCs/UCDs around NGC~1399. This sample is a compilation of many different studies \citep[][and Puzia \& Hilker (private communication)]{hilker99, drinkwater00, mieske02, mieske04, bergond07, mieske08, hilker07, firth07, gregg09, schuberth10, chilingarian11}. Since GCs/UCDs have roughly the same age, they probably have a comparable $M/L$-ratio, for which reason these brightest objects are also among the most massive in the central Fornax galaxy cluster. Because of their brightness, they are particularly suitable for spectroscopic analyses. The membership of these objects is confirmed by measurements of their radial velocity, extracted from their spectra. For this
reason, the spectroscopic sample offers very reliable number counts at the high-mass end of the GC/UCD mass function.

\paragraph{Photometric sample ('phot' sample):}
Our second sample contains 6268 objects, mostly GCs, around NGC~1399 and other central Fornax cluster galaxies based on HST/ACS observations, allowing resolved images of the GCs, reported by \citet{jordan07}. Since much fainter objects can be detected through photometry, GCs of much lower masses can be identified. Thus, as a result of the large number of objects and the lower mass limit, the photometric sample is  statistically more reliable than the spectroscopic sample, particularly in the intermediate- and low-mass regime.

\begin{figure}[t]
\includegraphics[angle=-90, width=0.488\textwidth]{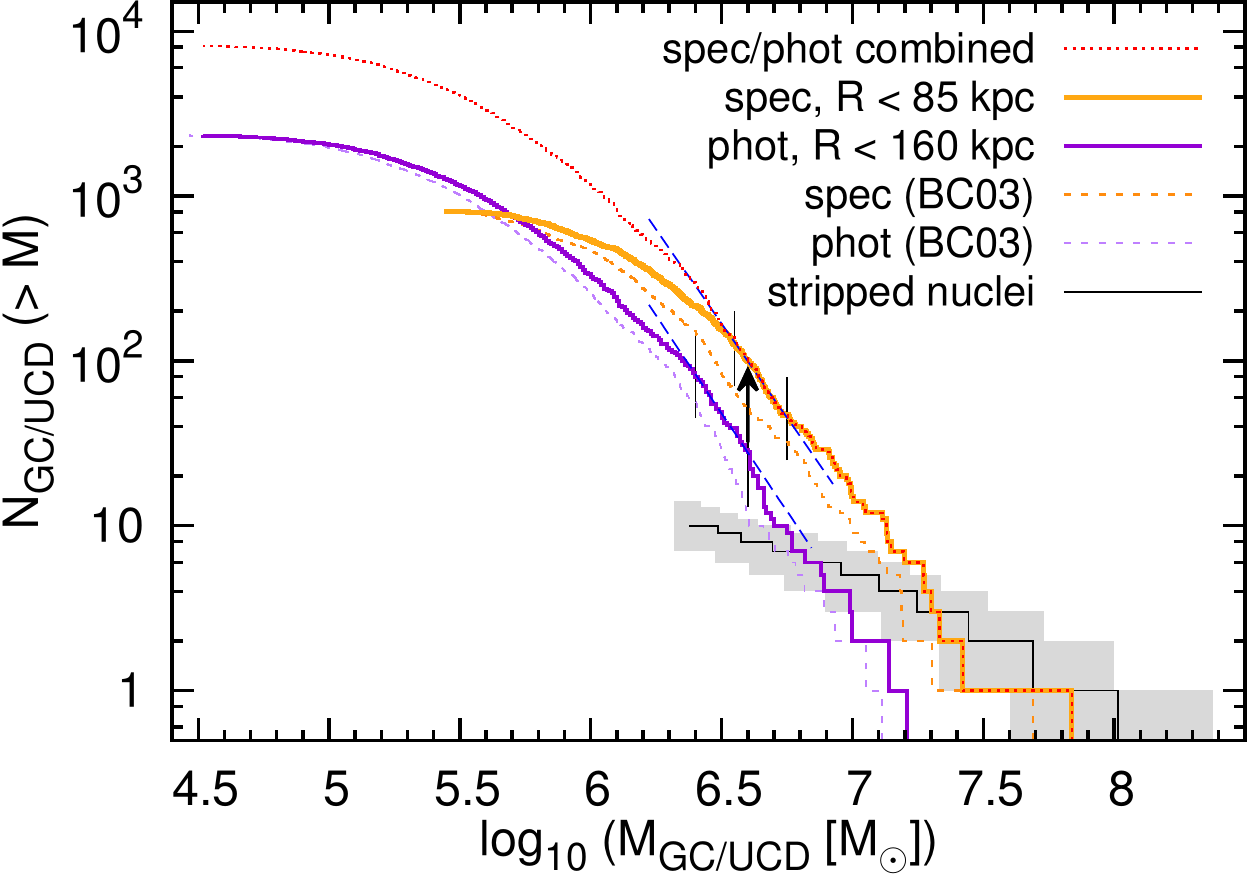}
\caption{Cumulative mass function of the spectroscopic (orange continuous line, spec sample) and the photometric sample (purple continuous line, phot sample). In both cases, the masses were determined from SSP models by \citet{maraston05}. The scaling with the factor 3.5 was applied at $\log_{10} (M/M_{\odot}) = 6.6$ (vertical arrow) where the slopes (dashed blue lines) of the cumulative functions are almost the same, resulting in the combined cumulative GC/UCD mass function (red dotted line). For comparison, the dashed thin orange and purple lines show the cumulative mass function of the spectroscopic and the photometric sample, respectively, where the masses have been calculated based on \citet{bc03}. The black line represents the cumulative distribution of stripped nuclei as determined by \citet{pfeffer14}, while the standard deviation area is colored in gray (see Sect.~\ref{sect_decomposition}).}
\label{fig_used_scaling}
\end{figure}

The masses of the GCs/UCDs were determined as follows: for the photometric sample, the $g$- and $z$-band photometry was converted into $M_V$ and $V-I$ based on the calibration by \citet{peng06XI}. For the spectroscopic sample, $V$, $V-I$, or $C-R$ were used to obtain the latter quantities. The individual GC/UCD masses were then determined using the obtained $M_V$ and $V-I$, a Kroupa IMF \citep{kroupa01} and a mass-to-light ratio, $M/L_V$, derived from models by \citet{maraston05}, where for each GC/UCD a 13~Gyr old simple stellar population (SSP) was assumed \citep{misgeld_hilker11, mieske13, puzia14}. For comparison, the individual GC/UCD masses were also determined based on $M/L_V$ obtained from models by \citet{bc03}, also assuming a 13~Gyr old SSP and using the same $M_V$ and $V-I$, but a Chabrier IMF \citep{chabrier03}. For our analysis, we used the masses as calculated from \citet{maraston05} and show below how much these masses differ from those determined based on \citet{bc03}.

Since our investigations are focused on the surroundings of NGC~1399, we have to apply distance cuts: The objects in the spectroscopic sample are concentrated around NGC~1399, but their distribution becomes patchy beyond 85~kpc. To obtain a spatially homogeneous sample, we included only objects within that radius, which led to a selection of 801 objects. On the other hand, the photometric sample comprises objects that are located around other galaxies in the Fornax galaxy cluster more than 1~Mpc away from NGC~1399 and thus not associated with it. To obtain a statistically representative sample, we included only objects that are within a radius of 160~kpc around NGC~1399 since this region is dominated by the central giant elliptical. This led to a selection of 2326 objects in the photometric sample.

Applying these distance cuts, we show the resulting cumulative GC/UCD mass functions where the GC/UCD masses were determined from models by \citet{maraston05} with a Kroupa IMF in Fig.~\ref{fig_used_scaling} for the spectroscopic sample in orange and for the photometric sample in purple. In the same colors but with thin dashed lines, the same mass distributions are shown for the masses determined based on \citet{bc03} with a Chabrier IMF. Figure~\ref{fig_used_scaling} shows that the models by \citet{bc03} predict somewhat lower masses than models by \citet{maraston05}: On average, the objects are 20\% and 15\% less massive in the spectroscopic and photometric sample, respectively. We conclude that the uncertainty in the mass determination is around 20\% on average even if there are cases where the masses deviate by up to one third.

To obtain an accurate number distribution across the whole mass range, the two samples have to be combined. This needs to be
done in such a way that each sample covers the mass regimes where it is more reliable. Consequently, the spectroscopic sample should determine the high-mass end of the GC/UCD mass function, while the photometric sample should define the shape of the GC/UCD mass function at the intermediate- and low-mass range. To achieve this, it is necessary to scale the photometric sample such that it matches the spectroscopic sample. The scaling must be done in a mass region that is not too high, where the photometric sample is inaccurate, but not too low where the spectroscopic sample becomes incomplete. We determined this overlapping region by demanding the same slope for both mass functions. As indicated by the vertical lines in Fig.~\ref{fig_used_scaling}, we selected the interval [6.55, 6.75] for the spectroscopic sample where the corresponding cumulative mass function has a slope of 3.3 and the interval [6.4, 6.6] for the photometric sample with a similar slope of 3.4. These two slopes are included in Fig.~\ref{fig_used_scaling} with blue dashed lines. At $\log_{10} (M/M_{\odot}) = 6.6$, there are 100 objects in the spectroscopic sample, while the photometric sample contains 29 GCs, leading to a scaling factor of 100/29 $\approx$ 3.5 with an error of about 0.7 when only considering Poisson noise. When the scaling point is chosen at a lower mass of $\log_{10} (M/M_{\odot}) =$ 6.55, 6.5, or 6.45 we obtain a scaling factor of 3.2, 3.1, or 2.9, respectively, because the spectroscopic mass function flattens toward lower masses. Choosing the scaling point at $\log_{10} (M/M_{\odot}) =$ 6.65, 6.7, or 6.75 gives a scaling factor of 4.5, 5.2, or 4.7, respectively, because the mass function of the photometric sample falls off steeply above $\log_{10} (M/M_{\odot}) = 6.6$. We therefore set the requirement that the scaling be done at a mass where the slopes of the spectroscopic and photometric mass functions are similar.

We combined the two samples as follows: The high-mass part above $\log_{10} (M/M_{\odot}) = 6.6$ (indicated by a vertical arrow in Fig.~\ref{fig_used_scaling}) is defined by the spectroscopic sample, while the intermediate- and the low-mass part emerges from shifting the photometric sample upward by the factor 3.5. The similar slope of the two samples in the overlap region leads to a smooth transition. The combined cumulative mass function of GC/UCD is represented by a red dotted line in Fig.~\ref{fig_used_scaling} and consists of 8143 objects. This is well within the estimate by \citet{gregg09} because we probe a smaller region. However, based on two different approaches, \citet{dirsch03} expected $6\,100 \pm 770$ and $6\,800 \pm 950$ objects, respectively, within 83 kpc (i.e.,~almost the same spatial region). Compared to their estimates, our combined sample includes slightly more objects.

\section{Correction of the observed GC/UCD sample} \label{sect_corrections}

To determine which distribution of SFRs has led to the formation of the observed GC/UCD sample, the individual mass of each GC/UCD at the time of its birth is required. Thus, it is necessary to correct the observed mass function for all aspects that resulted in a change of the individual GC/UCD masses. The main contributors are

\begin{enumerate}
\item mass loss due to stellar evolution,
\item mass loss due to dynamical evolution,
and\item elimination of objects that are not the result of an SC formation process.
\end{enumerate}

There are many elaborate models that treat the evolution of SCs as a function of various parameters, such as the orbit of the SCs, the concentration factor, and/or the Roche-lobe filling conditions \citep[e.g.,][]{lamers10, lamers13, alexander12, alexander14, brockamp14}. However, we decided to evaluate the corrections of the first two aspects based on the relatively simple model by \citet{lamers05}. Their model appears to be very suitable for our purposes since it requires only a handful of ingredients. This is important for us because the current knowledge about the combined GC/UCD sample is limited: We have a mass estimate for all objects based on their photometric properties. Additionally, spectra have been taken for the most luminous and therefore most massive objects, from which the radial velocity and some internal properties of these objects can be deduced. From photometry and/or spectroscopy, the metallicities and the ages of the GCs/UCDs are known. Typically, they lie between $-$2 dex and $+$0.5 dex, while the metal-poor and the metal-rich sub-populations peak at rougly $-$1.5 dex and $-$0.5 dex, respectively. Moreover, the projected 2D distance to the center of NGC~1399 is known for each object. 

However, external properties such as the 3D position in the galaxy cluster, the absolute velocity, or the parameters of the orbit are not known for any object of the sample. This means in turn that we cannot apply corrections regarding the internal and external dynamical evolution to each object individually without making assumptions, in particular for the lower mass tail of the combined GC/UCD sample, where number counts were extrapolated by scaling. Applying corrections based only on assumptions will lead to relatively large uncertainties that might prevent us from drawing reliable conclusions about the formation of the observed GC/UCD sample. To avoid this, we only applied corrections based on properties of the sample accessible to us.

The model by \citet{lamers05} allows us to correct for stellar evolution and the disruption of SCs in tidal fields, the two most important contributions regarding the mass loss of SCs. Using Eq.~(7) from \citet{lamers05}, we can calculate the initial mass of each SC, $M_{\mathrm{initial}}$, as a function of its present mass, $M_{\mathrm{now}}$, and its age, $t$:

   \begin{equation} \label{eq.corr}
      M_{\mathrm{initial}} (M_{\mathrm{now}}, t) = \left[ \left( \frac{M_{\mathrm{now}}}{M_{\odot}} \right)^{\gamma} + \frac{\gamma t}{t_0} \right]^{1/\gamma} \frac{1}{\mu_{\mathrm{ev}} (t)} .
   \end{equation}

\noindent Here, $\gamma = 0.62$ and $t_0$ can be expressed as $t_0 = (t_4/660)^{1/0.967}$ , where $t_4$ is the total disruption time of an SC with an initial mass of $10^4 M_{\odot}$. These parameters carry the information related to the mass loss due to tidal disruption for an SC of any mass. The mass loss due to stellar evolution is described by $\mu_{\mathrm{ev}} (t),$ which is the mass fraction of an SC with initial mass $M_{\mathrm{initial}}$ that is still bound at age $t$. With Eqs.~3 and 2 in \citet{lamers05}, $\mu_{\mathrm{ev}} (t)$ reads

   \begin{equation}
      \mu_{\mathrm{ev}} (t) = 1 - 10^{(\log (t) - a_{\mathrm{ev}})^{b_{\mathrm{ev}}} + c_{\mathrm{ev}} } ,
   \end{equation}

\noindent where $t > 12.5$ Myr must be fulfilled. Since we assumed the GCs/UCDs to have an age of $t = 13$~Gyr (Sect.~\ref{sect_sample}), the requirement is easily complied with. The parameters $a_{\mathrm{ev}}$, $b_{\mathrm{ev}}$, and $c_{\mathrm{ev}}$ characterize the mass loss by stellar evolution and depend on the metallicity, $Z$. The values of $a_{\mathrm{ev}}$, $b_{\mathrm{ev}}$, and $c_{\mathrm{ev}}$ can be found in Table~1 in \citet{lamers05}. We assumed the metallicity to be -0.8 dex on average for the whole GC/UCD sample, that means $10^{-0.8} Z_{\odot} = 0.00269$ with $Z_{\odot} = 0.017$ \citep[e.g.,][]{grevesse&sauval98} and $0.00212$ for a newer estimate $Z_{\odot} = 0.0134$ \citep[see also references therein]{asplund09}, respectively. According to these numbers, the closest match is $Z = 0.0040,$ for which the parameters read $a_{\mathrm{ev}} = 7.06$, $b_{\mathrm{ev}} = 0.26$, and $c_{\mathrm{ev}} = -1.80$.

\begin{table*}[tb]
\caption{Determination of the ambient density, $\rho_{\mathrm{amb}}$, based on three different mass models (R1 in Cols.~2-4, R2 in Cols.~5-7, and a10 in Cols.~8-10.) to estimate the lifetime of a $10^4~M_{\odot}$ SC, $t_4$, based on two different conversion relations (PZ in Cols.~3, 6, 9, and BM in Cols.~4, 7, 10) as a function of the radius $r$ (Col.~1).}
\label{tab_amb.den.t4}
\centering
\begin{tabular}{c|ccc|ccc|ccc}
\hline \hline
$r$     &       $\rho_{\mathrm{amb}}^{\mathrm{R1}}$             &       $t_4^{\mathrm{PZ}}$     &       $t_4^{\mathrm{BM}}$     &       $\rho_{\mathrm{amb}}^{\mathrm{R2}}$     &       $t_4^{\mathrm{PZ}}$     &       $t_4^{\mathrm{BM}}$     &       $\rho_{\mathrm{amb}}^{\mathrm{a10}}$    &       $t_4^{\mathrm{PZ}}$     &       $t_4^{\mathrm{BM}}$     \\
{[}kpc] &       [$M_{\odot} \mathrm{pc}^{-3}$]  &       [Gyr]   &       [Gyr]   &       [$M_{\odot} \mathrm{pc}^{-3}$]      &      [Gyr]    &       [Gyr]   &       [$M_{\odot} \mathrm{pc}^{-3}$]      &       [Gyr]   &       [Gyr]           \\
\hline
10      &       $2.95 \cdot 10^{-2}$    &       \cellcolor[gray]{0.78}$1.84$    &       \cellcolor[gray]{0.78}$4.62$    &       $2.26 \cdot 10^{-2}$  &       \cellcolor[gray]{0.78}$2.11$    &       \cellcolor[gray]{0.85}$5.29$    &       $1.79 \cdot 10^{-2}$  &       \cellcolor[gray]{0.78}$2.37$    &       \cellcolor[gray]{0.85}$5.94$    \\
20      &       $1.08 \cdot 10^{-2}$    &       \cellcolor[gray]{0.78}$3.04$    &       \cellcolor[gray]{0.85}$7.63$    &       $8.29 \cdot 10^{-3}$  &       \cellcolor[gray]{0.78}$3.47$    &       \cellcolor[gray]{0.85}$8.72$    &       $5.93 \cdot 10^{-3}$  &       \cellcolor[gray]{0.78}$4.11$    &       \cellcolor[gray]{0.93}$10.31$   \\
30      &       $5.53 \cdot 10^{-3}$    &       \cellcolor[gray]{0.78}$4.25$    &       \cellcolor[gray]{0.93}$10.68$   &       $4.23 \cdot 10^{-3}$  &       \cellcolor[gray]{0.78}$4.86$    &       \cellcolor[gray]{0.93}$12.21$   &       $2.82 \cdot 10^{-3}$  &       \cellcolor[gray]{0.85}$5.96$    &       $14.97$ \\
40      &       $3.28 \cdot 10^{-3}$    &       \cellcolor[gray]{0.85}$5.52$    &       $13.87$ &       $2.51 \cdot 10^{-3}$  &       \cellcolor[gray]{0.85}$6.32$    &       $15.86$ &       $1.58 \cdot 10^{-3}$  &       \cellcolor[gray]{0.85}$7.96$    &       $19.99$ \\
60      &       $1.46 \cdot 10^{-3}$    &       \cellcolor[gray]{0.85}$8.27$    &       $20.76$ &       $1.12 \cdot 10^{-3}$  &       \cellcolor[gray]{0.85}$9.45$    &       $23.74$ &       $6.52 \cdot 10^{-4}$  &       \cellcolor[gray]{0.93}$12.38$   &       $31.10$ \\
80      &       $7.86 \cdot 10^{-4}$    &       \cellcolor[gray]{0.93}$11.28$   &       $28.34$ &       $6.01 \cdot 10^{-4}$  &       \cellcolor[gray]{0.93}$12.90$   &       $32.40$ &       $3.33 \cdot 10^{-4}$  &       $17.34$ &       $43.55$ \\
100     &       $4.72 \cdot 10^{-4}$    &       $14.55$ &       $36.55$ &       $3.61 \cdot 10^{-4}$  &       $16.64$ &       $41.80$ &       $1.93 \cdot 10^{-4}$  &       $22.78$ &       $57.23$ \\
130     &       $2.52 \cdot 10^{-4}$    &       $19.91$ &       $50.01$ &       $1.93 \cdot 10^{-4}$  &       $22.77$ &       $57.19$ &       $9.89 \cdot 10^{-5}$  &       $31.79$ &       $79.86$ \\
160     &       $1.51 \cdot 10^{-4}$    &       $25.77$ &       $64.73$ &       $1.15 \cdot 10^{-4}$  &       $29.47$ &       $74.02$ &       $5.74 \cdot 10^{-5}$  &       $41.73$ &       $104.81$        \\
\hline
\end{tabular}
\end{table*}

The only ingredient that is not determined so far is $t_0$, which can be derived from $t_4$. Which would be a good estimate for the total disruption time of a $10^{4}~M_{\odot}$ SC around the giant elliptical NGC~1399? $t_4$ has been determined for M51, M33, the solar neighborhood, and the Small Magellanic Cloud by \citet[see their Table~1]{lamers05b}. Their values for $t_4$ vary between $10^{7.8}$~yr and $10^{9.9}$~yr. Their Table~1 and their Fig.~3 show that $t_4$ decreases with increasing ambient density, $\rho_{\mathrm{amb}}$, meaning that SCs are destroyed more easily in denser environments. This relationship is also found theoretically: Based on $N$-body simulations by \citet{pz98, pz02} (PZ) and \citet{baumgardt_makino03} (BM), \citet{lamers05b} showed in their Fig.~2 the dependence of $t_4$ on $\rho_{\mathrm{amb}}$, and two predicted lines that pass through the data points of each set of simulations. These two relations can be approximated by

   \begin{align}
\mathrm{PZ}: \quad \log(t_4) &= -0.5 \log(\rho_{\mathrm{amb}}) + 8.5 \label{eq_pz}\\
\mathrm{BM}: \quad \log(t_4) &= -0.5 \log(\rho_{\mathrm{amb}}) + 8.9 \label{eq_bm}.
   \end{align}

\noindent We used this correlation to determine the parameter $t_4$, which requires the ambient density profile around NGC~1399. Moreover, the above two relations also allow us to estimate the uncertainty of $t_4$.

To assess the ambient density around NGC~1399, we used Fig.~22 of \citet{schuberth10} where different approximations of the cumulative mass distribution of NGC~1399 as a function of the radius are shown. We selected the three models labeled R1, R2, and a10 to investigate the effect on $t_4$. These models were selected because they represent the full range of possible solutions to the observed mass distribution \citep[see Fig.~22 in][]{schuberth10}. Their mass profiles emerge from the following model parameters:

   \begin{align}
\mathrm{R1}:  \quad \rho_{\mathrm{s}} &= 0.0085~M_{\odot} \mathrm{pc}^{-3}, \quad r_{\mathrm{s}} = 50~\mathrm{kpc} \label{eq_r1}, \\
\mathrm{R2}:  \quad \rho_{\mathrm{s}} &= 0.0065~M_{\odot} \mathrm{pc}^{-3}, \quad r_{\mathrm{s}} = 50~\mathrm{kpc} \label{eq_r2}, \\
\mathrm{a10}: \quad \rho_{\mathrm{s}} &= 0.0088~M_{\odot} \mathrm{pc}^{-3}, \quad r_{\mathrm{s}} = 34~\mathrm{kpc} \label{eq_a10},
   \end{align}

\noindent where $r_{\mathrm{s}}$ is a core radius and $\rho_{\mathrm{s}}$ the central density. The two models R1 and R2 were taken from \citet{richtler08}. For all these models, the corresponding profile of the ambient density as a function of the radius is expressed by Eq.~(10) in \citet{richtler04}, where $\zeta = 1$ was used (cf.~Eqs.~(11) and (12) in \citealt{richtler04} with Eq.~(3) in \citealt{richtler08}):

   \begin{equation}
\rho_{\mathrm{amb}} = \frac{\rho_{\mathrm{s}}}{(r/r_{\mathrm{s}}) (1 + r/r_{\mathrm{s}})^2} .
   \end{equation}

\noindent Using this equation, we calculated the ambient densities at different radii for the three models R1, R2, and a10 (Eqs.~(\ref{eq_r1}) - (\ref{eq_a10})) and converted the resulting ambient densities, $\rho_{\mathrm{amb}}$, into lifetimes of a $10^4~M_{\odot}$ SC, $t_4$, according to the two above relations (Eqs.~(\ref{eq_pz}) and (\ref{eq_bm})). All results can be found in Table~\ref{tab_amb.den.t4}. The underlaid gray shading of individual entries in the table shows how strongly that particular value for $t_4$ would influence the correction of the observed GC/UCD mass function: the stronger the effect, the darker the color (dark gray: $t_4 <$ 5~Gyr, medium gray: 5~Gyr $< t_4 <$ 10~Gyr, light gray: 10~Gyr $< t_4 < t_{\mathrm{Hubble}}$).

\begin{figure}[b]
\includegraphics[angle=-90, width=0.488\textwidth]{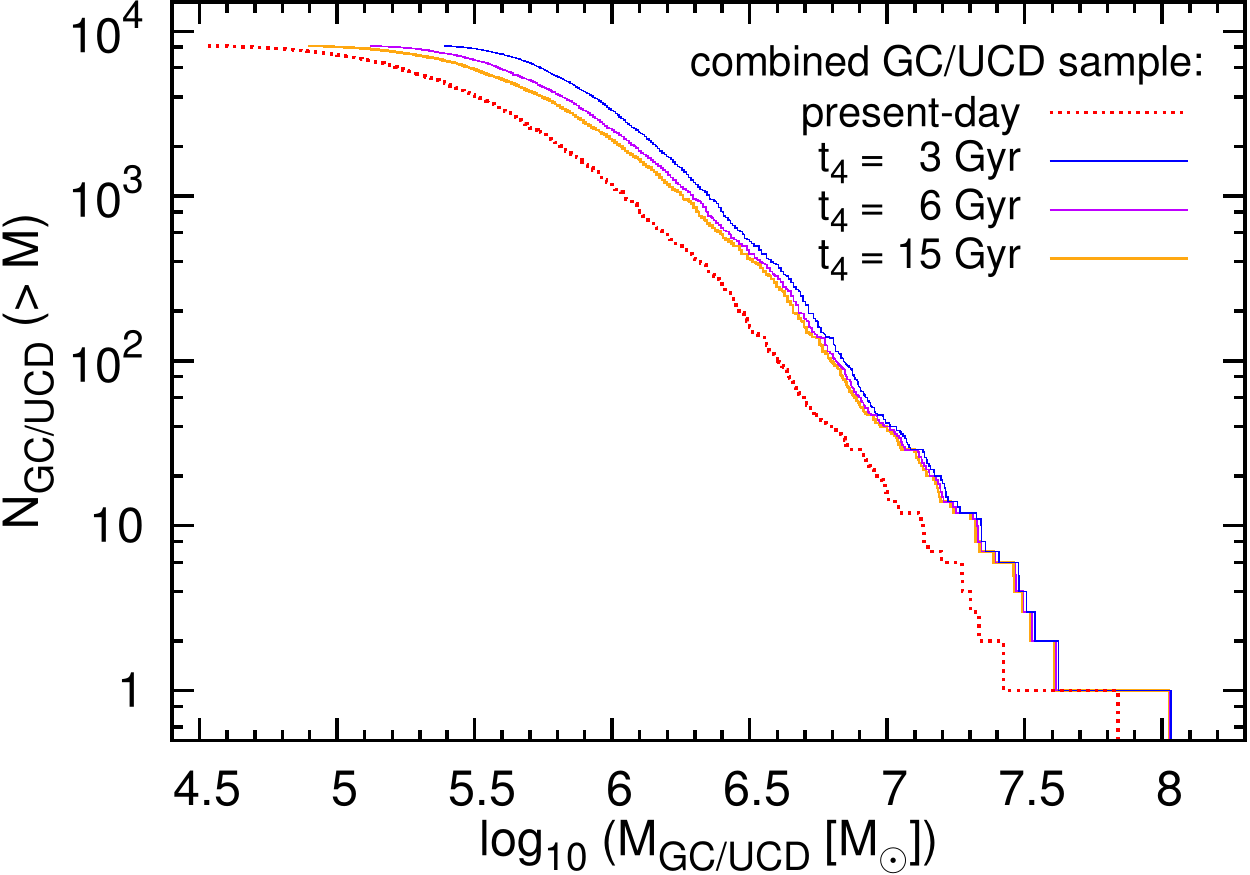}
\caption{Present-day (red dashed line) and natal cumulative mass functions (continuous lines) of the combined GC/UCD sample. The latter resulted from the former based on Eq.~\ref{eq.corr} with different values for the lifetime of a $10^4~M_{\odot}$ SC, $t_4$ (blue: $t_4 = 3$~Gyr, purple: $t_4 = 6$~Gyr, orange: $t_4 = 15$~Gyr). All other parameters are described in Sect.~\ref{sect_corrections}.}
\label{fig_corr_mass_func}
\end{figure}

Table~\ref{tab_amb.den.t4} lists the resulting values for $t_4$ , which shows that it depends much on the radius, the model used
for the ambient density, and its conversion to $t_4$. First, $t_4$ increases with increasing radius, which is expected since the ambient density decreases at the same time. Second, compared to $t_4$ values of the mass model R2, the corresponding $t_4$ values of the mass model R1 are lower, while the $t_4$ values of the mass model a10 are higher. This is also expected because the mass in model R1 increases more strongly with radius than model R2, while the increase of mass with radius is weaker for model a10 (cf.~Fig.~22 in \citealt{schuberth10}). The $t_4$ values of the different models reflect that the ambient density changes with radius in the same way as the mass does. Third, the conversion relation BM generally leads to longer lifetimes of $10^4~M_{\odot}$ SCs than the relation PZ. This is caused by the larger second term in the equation (cf.~Eqs.~(\ref{eq_pz}) and (\ref{eq_bm})). 

Interestingly, the conversion relations influence the resulting $t_4$ values much more strongly than the choice of the mass model: The relation BM gives values for $t_4$ more than twice as high as those from the relation PZ, while the differences in $t_4$ for the three mass models R1, R2, and a10 are on a 10\% level that slightly increases with radius, $r$. Thus, the primary influence determining the value of $t_4$ is the conversion relation and
not the mass model. The shortest survival time of a $10^4~M_{\odot}$ SC of about 2~Gyr is obtained near the center of NGC~1399 with mass model R1 and the conversion relation PZ, while a similar SC can outlast several Hubble times in the outskirts of NGC~1399.

Apparently, there is no one single value for $t_4$ that comprises all information about the dynamical evolution of the observed GCs/UCDs that have a variety of masses and distances to the center of NGC~1399. We therefore used different approximations for $t_4$ to see how much our analysis depend on that parameter. The first value we assumed was $t_4 = 15$~Gyr, which is somewhat longer than the Hubble time. In this case, the correction term (Eq.~(\ref{eq.corr})) is dominated by mass loss due to stellar evolution, making the mass loss due to dynamical evolution negligible for all our GCs/UCDs. Guided by the $t_4$ values based on the conversion relation by PZ in Table~\ref{tab_amb.den.t4}, we assumed in two comparison cases the lifetime of a $10^4~M_{\odot}$ SC to be $t_4 = 6$~Gyr and $t_4 = 3$~Gyr. In particular, $t_4 = 3$~Gyr will allow us to determine how strongly this parameter influences our analysis. In this case, the strongest effect on the combined GC/UCD mass function is expected to occur at its low-mass end.

The last aspect listed in the enumeration in the beginning of this section, the elimination of objects that did not form in a typical SC formation process such as stripped nuclei of dwarf galaxies or merged super SCs, is also a challenge. As a first approach, we assumed that all objects in our combined GC/UCD sample are genuine SCs, but we also investigated two alternatives in Sect.~\ref{sect_decomposition}.

Now, all ingredients for the mass correction are available: All parameters were chosen as described above, while for $t_4$ the values 3, 6, and 15~Gyr were assumed. The present mass of each GC/UCD in our combined sample (Fig.~\ref{fig_used_scaling}) is inserted into Eq.~(\ref{eq.corr}) to determine its initial mass. The corrected cumulative mass functions can be viewed in Fig.~\ref{fig_corr_mass_func}, where they are drawn by blue, purple, and orange continuous lines for the $t_4$ values 3, 6, and 15~Gyr, respectively, while the present-day mass function is indicated by a red dotted line. The corrected mass functions represent the mass distributions of the GCs/UCDs at their birth and were used as the starting point to determine their formation history.

As compared to the present-day mass function, the corrected mass functions are generally shifted to higher masses since SCs only lose but do not gain mass in the course of time, although under some circumstances further mass growth is possible \citep{bekki09, pflamm-altenburg09}. Apparently, the shift at the high-mass end is almost the same for all three values of $t_4$: The reason is that essentially all mass loss is caused by stellar evolution, which depends on the mass itself, while the tidal field, and thus $t_4$, has almost no influence on high-mass SCs. On the other hand, the shift at the low-mass end differs significantly among the three mass functions: As compared to high-mass SCs, SCs of lower masses are much more strongly exposed to the tidal field and thus lose a higher fraction of their mass, while the relative amount of mass lost due to stellar evolution remains the same. Thus, when correcting for mass loss, those SCs gain more mass relative to their present mass than high-mass SCs. This results in a steepening of the mass function at the low-mass end with decreasing $t_4$.

We note that this correction cannot tell how many SCs have been destroyed in the course of time: we can only trace back the mass loss of GCs that still exist, but do not have any indication of how many GCs have been destroyed over the same period of time. The number of destroyed SCs should increase with decreasing mass and decreasing $t_4$. On the other hand, the lifetime of any SC must have been longer in the past since the mass of the central elliptical NGC~1399 and the surrounding Fornax galaxy cluster increased to its present-day value, leading to longer survival times in the past. It is not obvious to which extent these two effects might counteract each other. Nevertheless, since destroyed GCs are not accounted for, this implies that in particular the number of low-mass GCs is probably underestimated.

We assume that the real natal mass function of the GC/UCD sample lies somewhat above the cumulative mass function described by $t_4 = 15$~Gyr. The latter represents the case of minimum requirement where all GC/UCD are corrected for stellar evolution while the influence of the tidal field becomes negligible. At least toward higher masses, the mass function with $t_4 = 3$~Gyr can be interpreted as a rough upper limit: for this mass function it is assumed that all objects have such a low $t_4$ value. This cleary is an overestimate since only the innermost objects have low $t_4$ values, and these objects only constitute a fraction of the whole sample. Still, at higher masses, the mass function is probably relatively accurate since the influence of $t_4$ is marginal and the complete dissolution of high-mass GCs/UCD is unlikely. However, in particular toward smaller masses, even the mass function with $t_4 = 3$~Gyr is probably an underestimate since destroyed GCs are not accounted for and low-mass GCs are particulary susceptible to dissolution \citep[e.g.,][]{fall_rees77,okazaki95, elmegreen10}. Bearing this in mind, we use all three corrected GC/UCD mass functions from Fig.~\ref{fig_corr_mass_func} in the following section to determine the variation this introduces in the distribution of necessary SFRs.

\begin{figure}[t]
\includegraphics[angle=-90, width=0.488\textwidth]{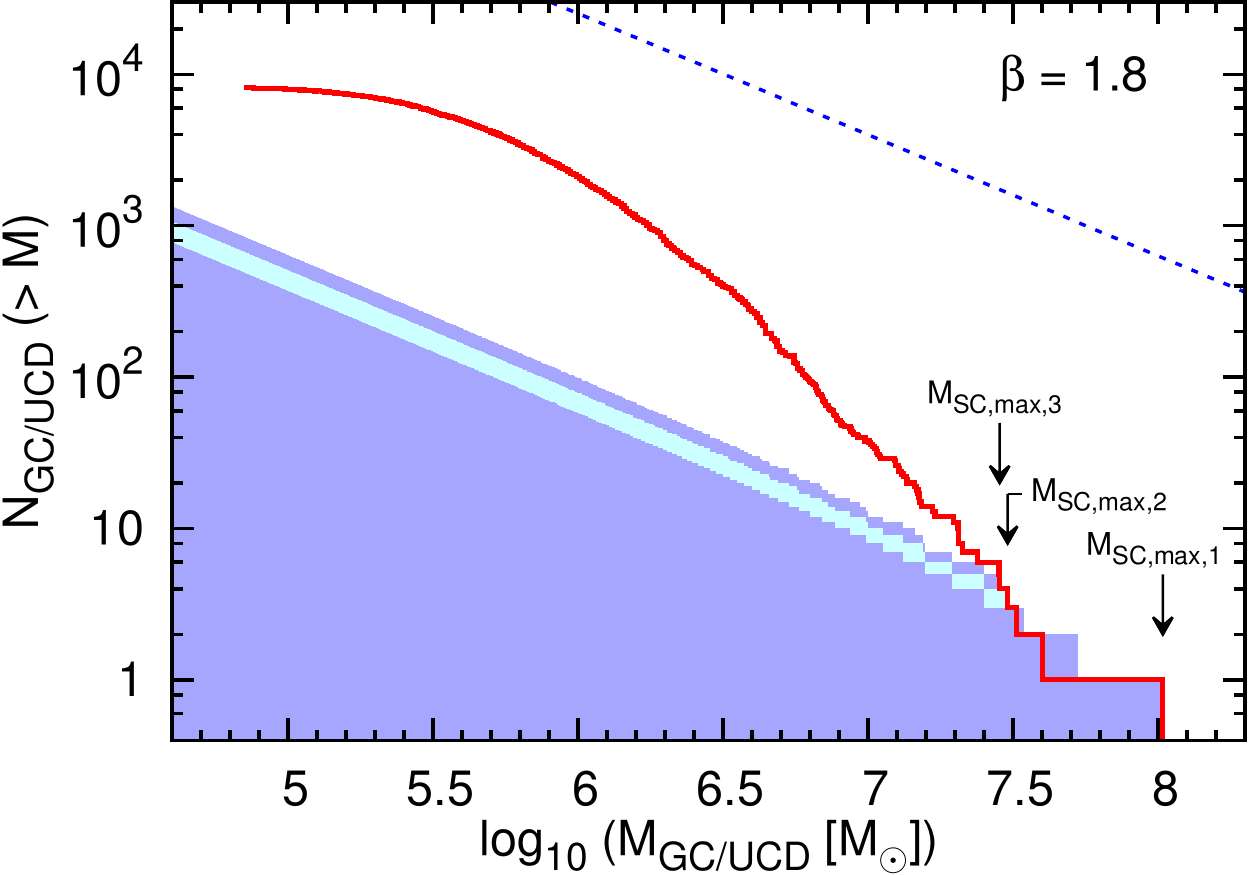}
\caption{Decomposition of  the observed GC/UCD cumulative mass distribution corrected for mass loss (see Sect.~\ref{sect_corrections}; red continuous line) into individual SC populations (colored areas). Here, the first three populations for an ECMF with $\beta = 1.8$ are shown. The most massive object of each population is indicated.}
\label{fig_decompo_sketch}
\end{figure}

\begin{figure*}[t]
\includegraphics[angle=-90, width=\textwidth]{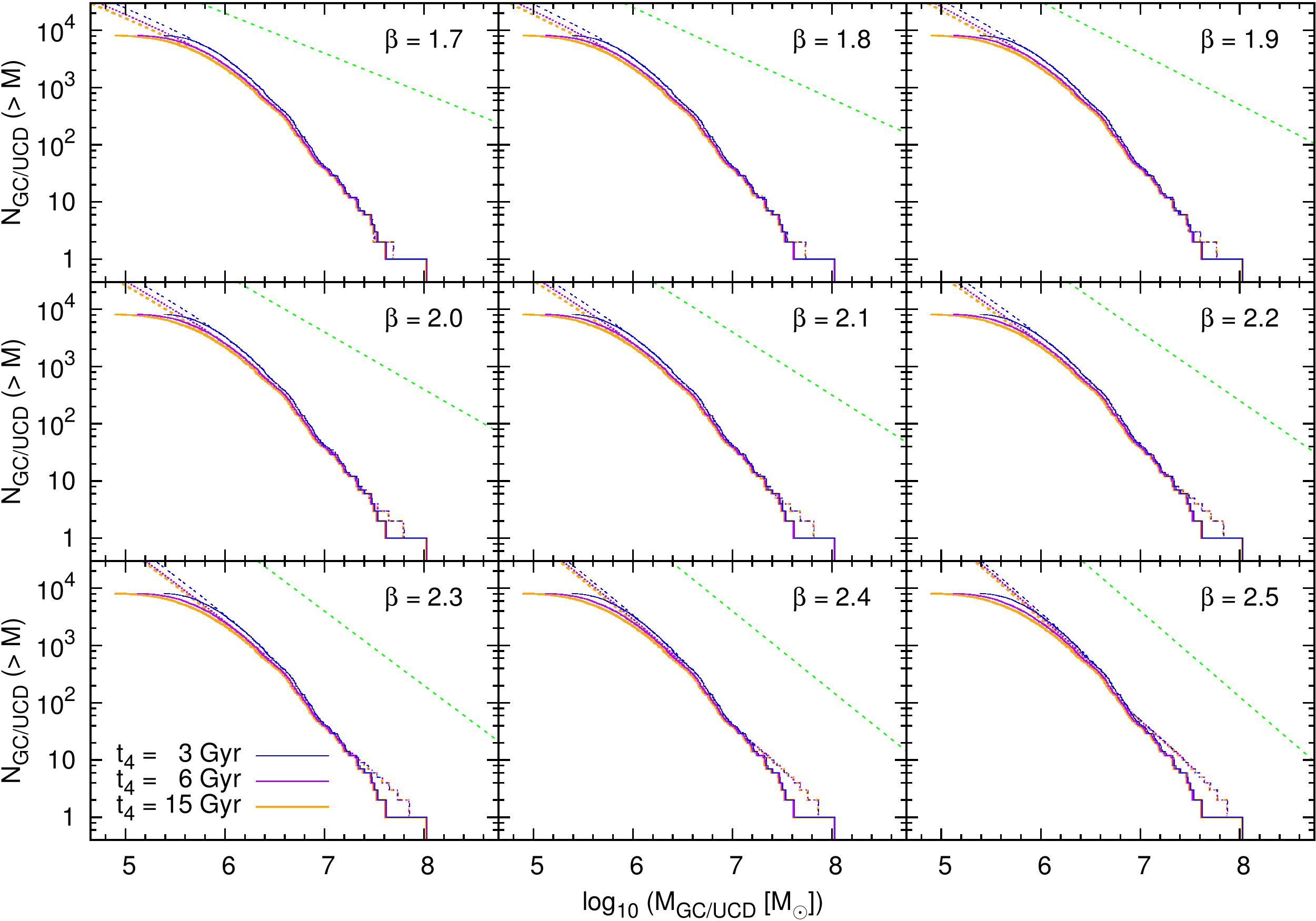}
\caption{Natal GC/UCD cumulative mass functions (continuous lines) together with the replicated ones (short dashed lines) as a function of $\beta$ (green dashed lines). The color scheme is the same as in Fig.~\ref{fig_corr_mass_func}.}
\label{fig_mass}
\end{figure*}

\begin{figure*}[t]
\includegraphics[angle=-90, width=\textwidth]{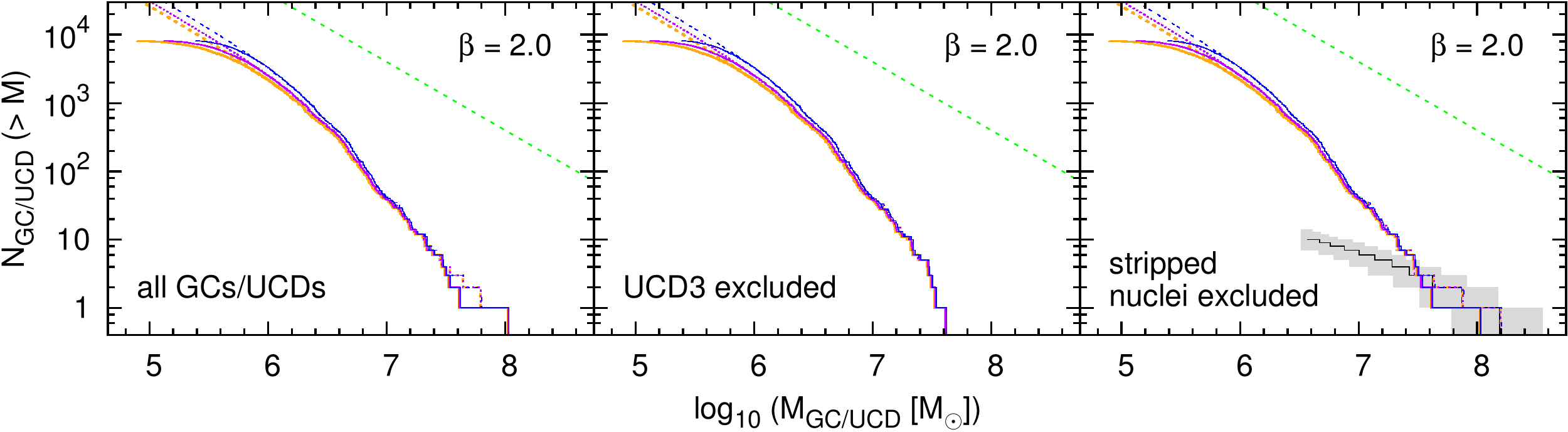}
\caption{Same as Fig.~\ref{fig_mass}, but for $\beta = 2.0$ (green dashed lines) for the standard approach (left panel), after excluding UCD3, the most massive object in our sample (middle panel), and using the stripped nuclei sample to account for the most massive objects in our sample (right panel). The color scheme is the same as in Fig.~\ref{fig_corr_mass_func}.        }
\label{fig_mass_vgl}
\end{figure*}

\section{Replication and decomposition of the GC/UCD sample} \label{sect_decomposition}

After restoring the natal cumulative mass function of our combined GC/UCD sample (Fig.~\ref{fig_corr_mass_func}), we are now able to decompose it into separate SC populations, each described by the ECMF (Eq.~(\ref{ecmf})). We make use of the fact that the most massive SCs in the combined GC/UCD sample can only be formed during epochs with a high SFR, while low-mass SCs can be formed during any SC formation epoch (Eq.~\ref{sfr}). Thus, to determine which and how many formation epochs contributed to the overall GC/UCD mass function, the replication has to start at the high-mass end:

\begin{enumerate}
\item We select the (remaining) most massive SC, $M_{\mathrm{SC,max}}$, in our combined GC/UCD sample and convert it into the theoretical upper mass limit, $M_{\mathrm{max}}$, after rearranging Eq.~(\ref{M_eclmax}). \label{step1}
\item $M_{\mathrm{max}}$ determines the required SFR through Eq.~(\ref{sfr}). \label{step2}
\item The normalization constant $k$ (Eq.~(\ref{norm_k})) depends on $M_{\mathrm{max}}$ , which is known from step~\ref{step1}, and $\beta, $ which is varied in the range [1.7, 2.5]. With $\beta$, $k$, and $M_{\mathrm{max}}$, the ECMF (Eq.~(\ref{ecmf})) is fully determined. We note that $\beta$ also sets the length of one SC formation epoch, $\delta t$ (see Cols.~2 and 3 in Table~\ref{tab_statistics}). \label{step3}
\item The derived ECMF is integrated downward to calculate the individual SC masses, $M_{i}$, of the population formed in the same epoch as the most massive SC, $M_{\mathrm{SC,max}}$, selected in step~\ref{step1}. For this, the optimal sampling technique (Sect.~\ref{sub_sampling}) is used. Since the ECMF is a pure power law, the individual SC masses can be evaluated analytically \citep[see][]{schulz15}. \label{step4}
\item All generated SCs (also from previous runs, if existing) are accumulated and sorted according to their mass. Starting at the high-mass end, the masses of the most-massive, second most-massive, third most-massive, and so on, SC of the generated and observed distributions are compared pairwise. We accept a deviation of up to five percent. The comparison stops as soon as the mass of an SC in the generated sample is less massive than tolerated, as compared to its counterpart in the observed sample. This SC in the observed sample is regarded as the remaining most massive SC in the observed GC/UCD sample. Thereafter the loop restarts. \label{step5}
\end{enumerate}

A schematic plot of this procedure is shown in Fig.~\ref{fig_decompo_sketch} for an ECMF with $\beta = 1.8$: Our procedure generates a first population (lower blue colored area) of SCs based on the most massive SC, $M_{\mathrm{SC,max,}1}$, in the observed GC/UCD sample corrected for mass loss (see Sect.~\ref{sect_corrections}; red continuous line). From the mass $M_{\mathrm{SC,max,}2}$ on, the first generated sample starts to deviate from the observed distribution. Thus, this SC in the observed sample is regarded as the most massive of the second population (light blue colored area). The first and second population together start deviating from the observed sample at the SC with the mass $M_{\mathrm{SC,max,}3}$ , which is regarded as the most massive SC of the third population (upper blue colored area). This iteration process is repreated until all generated SCs together replicate the observed GC/UCD sample as precisely as possible. Based on these and all following $M_{\mathrm{SC,max,}i}$, the required SFR for each formation epoch is determined according to steps \ref{step1} and \ref{step2} in the above enumeration.

Our goal is to reproduce the overall shape of the GC/UCD mass distribution and not to generating exact matches between individual SCs. To achieve this, we allowed five percent tolerance as mentioned above. Moreover, when we compared the SCs in the generated and the observed sample pairwise from high to low masses, we kept track of the difference in mass for each SC pair. For instance, sometimes an SC in the generated sample was more massive than its counterpart in the observed sample. When this occurred, we checked whether the following SC pair could compensate for this  mass difference, and then we only accepted an SC pair with a mass difference above the mentioned tolerance.

The reason for the above approach, the tolerance and taking care of the mass difference, is threefold: First, in this way, the generation of an SC population is prevented if an SC of similar but slightly lower mass is available in the generated sample. This is done to avoid an overproduction of SCs that potentially do not have an equally massive counterpart in the observed GC/UCD sample since there is no way of excluding an SC from the generated sample once it is generated. Second, this ensures that the total mass in the generated and the observed sample are similar. This enabled us to obtain a match between the generated and observed sample in terms of the shape of the GC/UCD mass distribution and the total mass in it. Third, the five percent margin introduces some tolerance since our optimized sampling distributes the masses of SCs very smoothly. However, we also tested how the choice of a margin of five percent influences our analysis. For comparison, we assumed no tolerance at all (i.e., 0 \%) and a margin of twenty percent and compared the results; in the former case slighly more, in the second case slightly fewer SCs are generated. The influence is minor but is discussed in Sect.~\ref{discuss_assump}.

As a first approach, all GCs/UCDs in our combined sample were treated as being formed in an SC formation process. The replication of the observed GC/UCD sample is shown for $\beta$ in the range between 1.7 and 2.5 in Fig.~\ref{fig_mass}. The initial GC/UCD mass distributions are represented by continuous lines (blue: $t_4 = 3$~Gyr, purple: $t_4 = 6$~Gyr, orange: $t_4 = 15$~Gyr), while the corresponding generated distributions are drawn with short dashed lines of the same color. The green dashed lines indicate the underlying ECMF. The mass distributions generated with no margin and a twenty percent margin exhibit slightly more and fewer SCs, respectively. The mass functions themselves look essentially the same apart from the fact that they are slightly shifted upward and downward at the low-mass end, respectively, but have the same slope. To avoid overcrowding the figure, they are not shown because the difference is barely visible owing to the logarithmic scale.

Overall, and in particular for lower $\beta$, the above procedure works well: The generated distributions match the observed one nearly perfectly. Only at the low-mass end do the samples of generated SCs start to deviate from the observed GCs/UCDs sample because the distribution of GCs/UCDs flattens toward the lower mass end, while the underlying ECMF (green dashed lines in Fig.~\ref{fig_mass}) has the same slope throughout. As $\beta$ increases and the parental ECMFs steepen, the deviation at the low-mass end becomes more prominent. This deviation might be due to the fact that in the observed mass distribution, the survived GCs/UCDs were corrected for stellar and dynamical evolution, but the completely dissolved GCs/UCDs were not taken into account: If the masses of SCs are distributed according to a power law, as we assumed, it was shown in various studies that low-mass SCs are destroyed more efficiently than high-mass SCs \citep[e.g.,][]{fall_rees77,okazaki95, elmegreen10}. This leads to a Gaussian mass distribution, which is indeed observed around NGC~1399 \citep[see e.g.,][their Figs.~4 and 5]{hilker09}.

Another peculiarity appears in Fig.~\ref{fig_mass}: At the high-mass end, the SCs of the generated sample become more massive than their conterparts in the observed GC/UCD sample because $\beta$ increases, therefore more SCs are drawn from the underlying ECMF. In contrast, the most massive object in the observed sample is more than 2.5 times more massive than the second most massive object, which leads to a substantial mass gap in between.

All of the above findings are independent of the choice of the parameter $t_4$. The only difference between the generated mass distributions with a certain $\beta$ is that they are slightly shifted to higher masses in the same way as the low-mass end of the initial mass distributions.

The most massive object in our combined GC/UCD sample is UCD3, which \citet{frank11} found to be fully consistent with a massive GC when surveying its internal kinematics. However, it still remains a peculiar object: It has an effective radius of almost 90~pc \citep{evstigneeva07, hilker07, frank11}, which is much larger than the effective radii of typical GCs of about 3 to 5~pc \citep[e.g.,][]{drinkwater03, jordan05}. Moreover, its surface brightness profile is best fit with a two-component model \citep{drinkwater03, evstigneeva07}, meaning that UCD3 is described best by a core that is surrounded by a halo with effective radii of around 10 and 100~pc, respectively \citep{evstigneeva07}. Such a composition of a core and a halo could be interpreted as a not fully completed stripping process of a more extended object \citep{evstigneeva08}. However, the merged star cluster scenario is a possible formation channel as well \citep{fellhauer&kroupa05}. \citet{bruens12} emphasized that a core-halo surface brightness profile may also occur after the merging of SCs based on their simulations on the formation of super SCs in \citet{bruens11}.

Since UCD3 does not seem to be a typical GC, we tested how our analysis is influenced when it is removed from our combined GC/UCD sample. We kept everything else the same and reran our above method. There was only one difference compared to the previous run: the agreement at the high-mass end was much tighter. This finding is independent of $\beta$, for which reason we only show the resulting mass distributions for $\beta = 2.0$ in the middle panel of Fig.~\ref{fig_mass_vgl}, in comparison to the first approach where UCD3 was included (left panel of Fig.~\ref{fig_mass_vgl}). The overproduction of high-mass SCs clearly disappears completely. The situation at the low-mass end, meaning the dependence on $t_4$, remains the same as before.

There are two interpretations possible for this finding: First, if it is assumed that UCD3 is a genuine SC, then this would hint at a small $\beta$ since otherwise between one and three very massive SCs of similar mass should have formed in the same formation event (cf.~the overproduction of SCs at the high-mass end for large $\beta$ in Fig.~\ref{fig_mass}). However, this is not observed. It is unlikely that these objects exist because they would be among the brighest UCDs and thus hard to miss observationally. The mass gap between the most massive and the second most massive UCD (cf.~the high-mass end of the GC/UCD sample in Fig.~\ref{fig_mass}) together with the typical values for $\beta$ of around 2.0 to 2.3 \citep[e.g.,][]{zhang_fall99, lada_lada03, weidner04, mccrady_graham07, chandar11} indicate a second possibility: As already suggested by its internal properties, UCD3 cannot be classified as a normal GC that formed in a typical SC formation process. We regard this as the more probable possibility.

We discuss the question whether UCD3 might not be the only object that does not fall into the category 'genuine GC'. Unfortunately, no predictions have been made so far regarding the SC mass function for the merged star cluster scenario. However, for the dwarf galaxy threshing scenario, \citet{pfeffer16} estimated for a galaxy similar to NGC~1399 a possible number of objects originating from stripping a nucleated dwarf galaxy. Their expected cumulative distribution within 83~kpc around the central galaxy, this means similar to the distance cut we applied, is plotted in black, while the standard deviation area is colored in gray in Fig.~\ref{fig_used_scaling} (private communication). To be consistent, the same mass correction as described in Sect.~\ref{sect_corrections} was applied to the stripped nuclei sample by \citet{pfeffer16}.

\begin{figure*}[t]
\includegraphics[angle=-90, width=\textwidth]{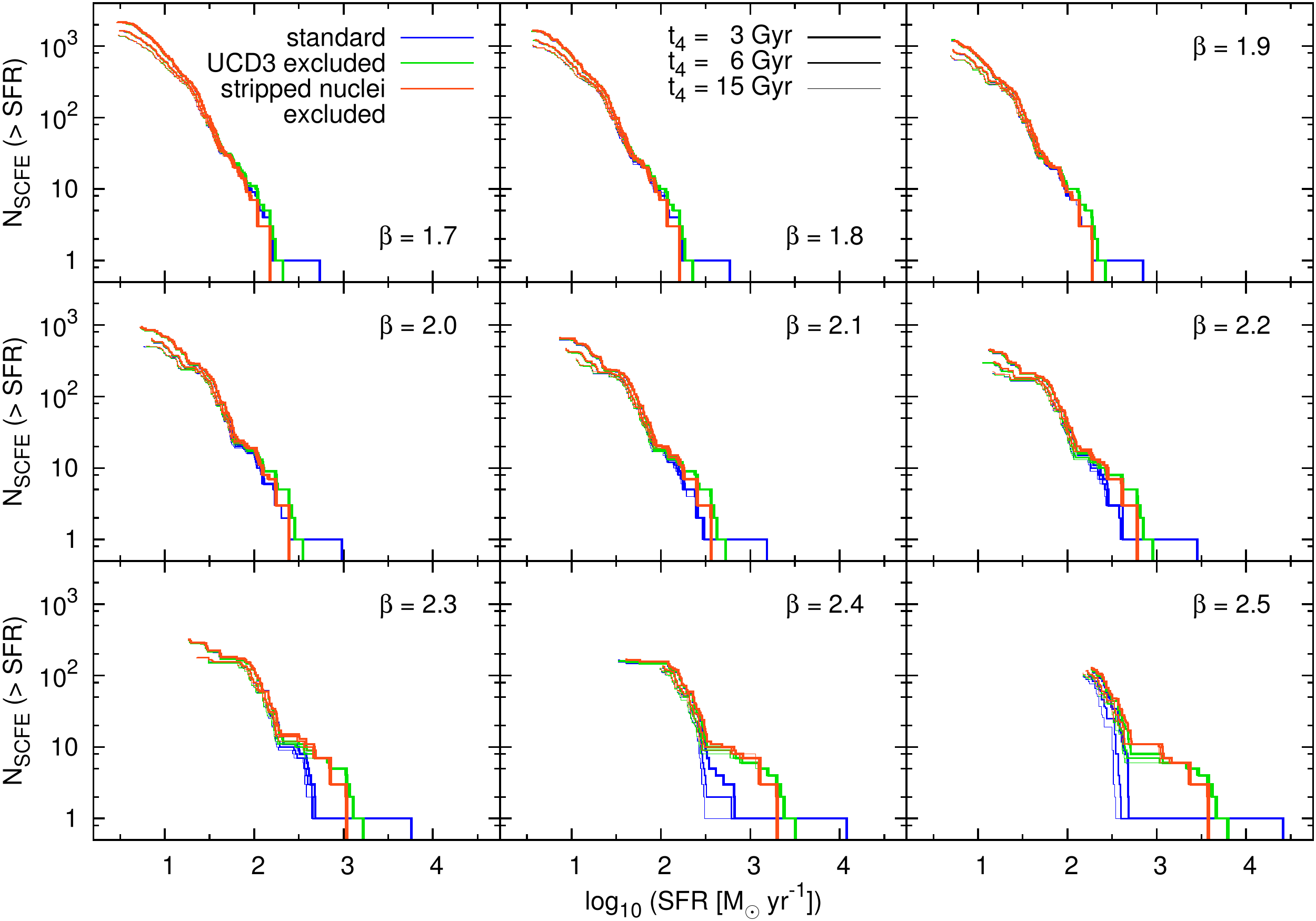}
\caption{Cumulative number of SC formation epochs (SCFEs) as a function of the SFR for the standard approach in blue, for the GC/UCD sample after excluding UCD3 in green, and for taking into account the stripped nuclei sample in red. The thickness of the lines depends on the value of the parameter $t_4$ (thick lines for $t_4 = 3$~Gyr, medium thick lines for $t_4 = 6$~Gyr, thin lines for $t_4 = 15$~Gyr).}
\label{fig_sfr}
\end{figure*}

In a third approach, we assumed that this stripped nuclei sample represents those objects in our combined GC/UCD sample (here, the most massive object, UCD3, is included again) that did not form in an SC formation process but are nuclei whose envelope was stripped away. Figure~\ref{fig_used_scaling} shows that the distribution of stripped nuclei accounts for the four most massive objects in the observed sample, which lie within the standard deviation area of the stripped nuclei sample. Consequently, our method needs to reproduce all remaining objects so that finally, the stripped nuclei sample together with the generated sample match the observed distribution of GCs/UCDs. For this, we started with the stripped nuclei sample and proceeded as before. The result of this third approach is shown in the right panel of Fig.~\ref{fig_mass_vgl} where the same color scheme is used.

The behavior at the low-mass end is essentially the same as in the two cases before: While the observed GC/UCD distribution flattens toward lower masses, the distribution of generated SC continues with the same slope as the underlying ECMF. In addition, the same shift to higher masses appears for shorter $t_4$. However, at the high-mass end the situation appears to be different from the first approach and similar to the previous one where UCD3 was excluded: Since the stripped nuclei sample accounts for the four most massive objects in the observed sample, there is no overproduction of SCs at the high-mass end. Instead, for all $\beta$, our algorithm accurately replicates the remaining GCs/UCDs distribution (right panel of Fig.~\ref{fig_mass_vgl}).

\section{Distribution of necessary SFRs} \label{sect_sfr.distr}

As mentioned in Sect.~\ref{sect_framework}, each SC population is characterized by its own individual stellar upper mass limit, $M_{\mathrm{max}}$. According to the SFR-$M_{\mathrm{max}}$ relation \citep{weidner04, randria13}, this mass limit can be translated into an SFR under which that SC population formed. Since we decomposed the observed GC/UCD mass function into individual SC populations, it is possible to determine the necessary SFR for each population: For each of the three approaches and for each of the initial GC/UCD mass distributions based on the three different $t_4$, $M_{\mathrm{max}}$ of each SC population is converted into an SFR using Eq.~(\ref{sfr}). This results in nine different cumulative SFR distributions that are plotted in Fig.~\ref{fig_sfr} for the different $\beta$. These SFR distributions show in a cumulative way how many GC/UCD formation events are necessary above a certain SFR. The color coding is as follows: The resulting SFR distributions for the standard approach, where all GCs/UCDs are kept, are marked in blue. The SFR distributions of the second approach where UCD3 was excluded are drawn in green, while those of the last approach where a sample of stripped nuclei was taken into account are shown in red. Moreover, the thickness of the lines representing the SFR distributions are varied depending on the $t_4$ value of the underlying GC/UCD birth mass function: We use a thick line for $t_4 = 3$~Gyr, a medium thick line for $t_4 = 6$~Gyr, and a thin line $t_4 = 15$~Gyr.

The resulting SFR distributions are remarkably similar for each $\beta,$  regardless of the different approaches and the $t_4$ value of the underlying GC/UCD birth mass function. The similarity is particularly striking toward smaller $\beta$. The details are as follows:

\begin{enumerate}
\item The resulting SFRs increase with $\beta$ and cover a range between $\log (\mathrm{SFR}) \approx$ 0.5 and 2.5 for $\beta = 1.7$, while for $\beta = 2.5$ the SFR range lies between $\log (\mathrm{SFR}) \approx$ 2.0 and 4.5. This is mostly independent of the choice of $t_4$ but depends on the treatment of the highest mass objects (see below). At the same time, the number of required GC/UCD formation events, $N_{\mathrm{SCFE}}$, decreases with $\beta$ from a few thousand events for $\beta = 1.7$ to roughly one hundred events for $\beta = 2.5$. Consequently, the main finding is that the higher $\beta$, the higher the SFRs and the fewer formation epochs, $N_{\mathrm{SCFE}}$, are needed to build up the entire GC/UCD sample (cf.~Cols.~4, 7, and 10 in Table~\ref{tab_statistics}).
\item As mentioned above, the SFR distributions become less similar with increasing $\beta$. In addition to the somewhat broader distribution of the SFR functions toward lower SFRs, the main difference is the distribution of the highest SFRs: In particular for $\beta = 2.5$, for the standard approach (blue lines) obviously only one GC/UCD formation event with a very high SFR is needed, while the other two approaches (green and red lines) require several formation events with a range of slightly lower SFRs. Even though the latter two are treating the objects at the high-mass end differently, their resulting SFR distributions are fairly similar, independent of $\beta$.
\item The highest peak SFRs are always obtained for the standard approach (all GCs/UCDs included, blue lines), the lowest peak SFRs always in the case when the stripped nuclei sample is taken into account (red lines). When only UCD3 is excluded (green lines), the peak SFRs are somewhat higher than taking into account the stripped nuclei sample. This is expected since lower SFRs are needed if the highest mass object(s) is excluded or is accounted for by the stripped nuclei sample, respectively.
\item At the low-SFR end, the SFR distributions develop into three different tails, depending on $t_4$ (thickness of the lines) but independent of how the high-mass end is treated (color). This is particularly visible for a not too high $\beta$. It shows that the number of GC/UCD formation events at the low-SFR end is solely defined by the shape of the low-mass end of the GC/UCD mass function, which itself is determined by $t_4$, and independent of the treatment of the high-mass end of the GC/UCD mass function. The number of formation events is highest for the smallest $t_4$ (3~Gyr, thick lines) while the smallest number of formation events is obtained for the highest $t_4$ (15~Gyr, thin lines). This simply represents the slighty higher/lower masses of the low-mass objects in the observed GC/UCD sample for lower/higher $t_4$ values, respectively. However, it needs to be considered that the SFR distributions particularly at the low-SFR end only represent a lower limit: the used mass functions are not corrected for destroyed GCs/UCDs, so that more formation epochs and/or higher SFRs might be necessary. Furthermore, the choice of a five-percent margin when replicating the observed GC/UCD mass function also influences the SFR distribution, but only at the low-SFR end since mainly the low-mass end of the mass function changes. For a larger/smaller margin, slightly fewer/more formation epochs and lower/higher SFRs are required. Since the difference is rather small, the resulting SFR distributions are not shown in Fig.~\ref{fig_sfr} for clarity (but see Sect.~\ref{discuss_assump} for a discussion).
\end{enumerate}

Given that the lowest and highest $\beta$ are only rarely observed and the standard approach, meaning the assumption that even the most massive UCDs formed as a single SC, is not well justified, we obtain peak SFRs of between $\log (\mathrm{SFR}) \approx$ 2.5 and 3.5 corresponding to values between roughly 300 and 3000 $M_{\odot} \mathrm{yr}^{-1}$. The question is how reasonable these SFRs are for the formation of a rich GC/UCD system like that observed around NGC~1399. Comparing this result to other studies in the literature, we discuss in Sect.~\ref{discuss_formation} in detail what the range of SFRs tells us about the formation of NGC~1399 itself.

\begin{table*}[tb]
\caption{Total number of SC formation epochs (SCFE), $N_{\mathrm{SCFE,tot}}$, the total SC formation time, $t_{\mathrm{SCFE,tot}}$, and the total stellar mass, $M_{\mathrm{tot}}$, formed during that time for the three different approaches as a function of $t_4$ (Col.~1) and $\beta$ (Col.~2). These quantities are listed for the standard approach in Cols.~4-6, for the case when the most massive object, UCD3, is excluded in Cols.~7-9, and for an assumed distribution of stripped nuclei in Cols.~10-12. For reference, the length of one SC formation epoch, $\delta t$, is listed in Col.~3 as determined in \citet{schulz15}.}
\label{tab_statistics}
\centering
\begin{tabular}{lcc||ccc|ccc|ccc}
\hline \hline
& & &   \multicolumn{3}{c}{standard approach}   &       \multicolumn{3}{c}{UCD3 excluded}       &       \multicolumn{3}{c}{stripped nuclei excluded}    \\
$t_4$   &       $\beta$ &       $\delta t$              &       $N_{\mathrm{SCFE,tot}}$ &       $t_{\mathrm{SCFE,tot}}$ &       $M_{\mathrm{tot}}$      &       $N_{\mathrm{SCFE,tot}}$ &       $t_{\mathrm{SCFE,tot}}$ &       $M_{\mathrm{tot}}$      &       $N_{\mathrm{SCFE,tot}}$ &       $t_{\mathrm{SCFE,tot}}$ &       $M_{\mathrm{tot}}$      \\
&       &       [Myr]   &               &       [Gyr]   &       [$10^{10}~M_{\odot}$]   &         &       [Gyr]   &       [$10^{10}~M_{\odot}$]   &            &       [Gyr]   &       [$10^{10}~M_{\odot}$]   \\
\hline
3~Gyr   &1.7    &       0.77    &       2168    &       1.66    &       1.98   &       2178    &       1.67    &       1.97   &       2186    &       1.67    &       1.95    \\
        &1.8    &       1.11    &       1639    &       1.82    &       2.52   &       1646    &       1.82    &       2.51   &       1655    &       1.83    &       2.50    \\
        &1.9    &       1.70    &       1211    &       2.06    &       3.56   &       1218    &       2.07    &       3.55   &       1226    &       2.09    &       3.53    \\
        &2.0    &       2.80    &       934     &       2.61    &       5.84   &       934     &       2.61    &       5.83   &       947     &       2.65    &       5.82    \\
        &2.1    &       4.94    &       671     &       3.31    &       11.11  &       669     &       3.30    &       11.08  &       688     &       3.40    &       11.12   \\
        &2.2    &       9.31    &       451     &       4.20    &       24.40  &       450     &       4.19    &       24.40  &       465     &       4.33    &       24.42   \\
        &2.3    &       18.57   &       323     &       6.00    &       61.94  &       321     &       5.96    &       61.86  &       332     &       6.17    &       61.83   \\
        &2.4    &       38.77   &       167     &       6.47    &       166.03 &       164     &       6.36    &       165.56 &       173     &       6.71    &       165.74  \\
        &2.5    &       83.77   &       121     &       10.14   &       516.62 &       126     &       10.56   &       522.28 &       133     &       11.14   &       519.83  \\
\hline
6~Gyr   &1.7    &       0.77    &       1671    &       1.28    &       1.55   &       1672    &       1.28    &       1.54   &       1683    &       1.29    &       1.52    \\
        &1.8    &       1.11    &       1225    &       1.36    &       1.96   &       1221    &       1.35    &       1.94   &       1239    &       1.37    &       1.93    \\
        &1.9    &       1.70    &       899     &       1.53    &       2.76   &       894     &       1.52    &       2.74   &       917     &       1.56    &       2.74    \\
        &2.0    &       2.80    &       651     &       1.82    &       4.48   &       652     &       1.82    &       4.47   &       661     &       1.85    &       4.45    \\
        &2.1    &       4.94    &       470     &       2.32    &       8.54   &       470     &       2.32    &       8.51   &       482     &       2.38    &       8.51    \\
        &2.2    &       9.31    &       287     &       2.67    &       18.55  &       294     &       2.74    &       18.62  &       309     &       2.88    &       18.65   \\
        &2.3    &       18.57   &       165     &       3.06    &       46.46  &       163     &       3.03    &       46.23  &       180     &       3.34    &       46.44   \\
        &2.4    &       38.77   &       135     &       5.23    &       140.82 &       134     &       5.19    &       139.86 &       139     &       5.39    &       138.43  \\
        &2.5    &       83.77   &       113     &       9.47    &       448.24 &       113     &       9.47    &       446.09 &       118     &       9.89    &       440.75  \\
\hline
15~Gyr  &1.7    &       0.77    &       1434   &       1.10   &       1.35   &       1436   &       1.10   &       1.34   &       1454   &       1.11   &       1.32   \\
        &1.8    &       1.11    &       1025   &       1.14   &       1.69   &       1025   &       1.14   &       1.68   &       1036   &       1.15   &       1.66   \\
        &1.9    &       1.70    &       741    &       1.26   &       2.38   &       746    &       1.27   &       2.37   &       753    &       1.28   &       2.35   \\
        &2.0    &       2.80    &       507    &       1.42   &       3.81   &       503    &       1.41   &       3.78   &       515    &       1.44   &       3.77   \\
        &2.1    &       4.94    &       339    &       1.67   &       7.19   &       336    &       1.66   &       7.16   &       347    &       1.71   &       7.16   \\
        &2.2    &       9.31    &       212    &       1.97   &       15.87  &       209    &       1.95   &       15.85  &       224    &       2.08   &       15.87  \\
        &2.3    &       18.57   &       146    &       2.71   &       42.34  &       144    &       2.67   &       41.96  &       149    &       2.77   &       41.60  \\
        &2.4    &       38.77   &       125    &       4.85   &       127.39 &       126    &       4.88   &       127.29 &       129    &       5.00   &       125.08 \\
        &2.5    &       83.77   &       102    &       8.54   &       403.35 &       105    &       8.80   &       404.42 &       110    &       9.22   &       400.12 \\
\hline
\end{tabular}
\end{table*}

\section{Discussion} \label{sect_discussion}

For our investigation, we made several assumptions that we review in the first part of this section. In the second part we continue with a discussion of our results while in the third part we focus on the formation of NGC~1399 and its GC/UCD system.

\subsection{Assumptions} \label{discuss_assump}

Our main assumption is that it is possible to decompose the observed GC/UCD sample into individual SC populations that formed at the same time out of the same molecular cloud. This approach can be applied to our data because it is known that the GCs/UCDs are of similar age. However, we cannot prove that those GCs/UCDs that are assumed to form a population indeed formed together. Our sample is comprehensive, therefore this assumption is not too strong because particularly toward the low-mass end, there are many GCs with similar masses, making them exchangeable. It should be noted that our approach is of a statistical nature and not a deterministic analysis.

Even though we took into account that a part of the UCDs might not originate from an SC formation process, it is not entirely clear whether the bulk of UCDs are compatible with being massive GCs. According to \citet{gregg09}, the UCDs in the Fornax galaxy cluster form a dynamically distinct population compared to the GC system, with a higher mean velocity and a lower velocity dispersion \citep[see also][]{mieske04}. This might indicate a different formation process but does not imply in general that UCDs are not SCs since the most massive SCs have probably formed in the most intense star-forming region and may therefore have a different kinematic signature. Moreover, it has been suggested that only the fainter and less massive UCDs could be genuine GCs while the brightest and most massive ones might have formed by tidal threshing \citep{mieske04, chilingarian11}. However, \citet{mieske12} restricted the fraction of tidally stripped dwarfs to not more than 50\% of UCDs with masses above $2 \cdot 10^6~M_{\odot}$ based on statistical considerations. \citet{pfeffer14} expected roughly 12 and 20~stripped within 83~kpc and 300~kpc around NGC~1399, whereas almost 150 and 200~GCs/UCDs are observed, respectively (see their Table~2). The contribution for lower masses becomes insignificant. This is well within the constraints set by \citet{mieske12}. More recently, \citet{pfeffer16} estimated that stripped nuclei account for around 40\% of the GCs/UCDs above $10^7~M_{\odot}$ , while for masses between $10^6~M_{\odot}$ and $10^7~M_{\odot}$ the contribution drops to about 2.5\%. As the authors emphasized, this implies that not all of the objects observed at the high-mass end of the GC/UCD mass function can be explained by tidally stripped dwarf galaxies.

If most of the UCDs are indeed of SC origin and not threshed dwarfs, it is not known whether they might have formed in the merged SC scenario \citep[e.g.,][]{fellhauer&kroupa02, bruens12}. At the time of formation, the newly born massive SCs were most likely embedded in a high-density environment, allowing a part of these SCs to merge into more massive super SCs. In this case, the masses of the pristine SCs are distributed according to the ECMF (Eq.~(\ref{ecmf})), but the masses of the final SC population might be distributed fairly differently. Consequently, the observed GC/UCD mass function cannot be decomposed into SC populations that are described by the ECMF.

It is not clear whether and how the formation of SCs changes in case of very high SFRs as derived by us. The stellar IMF may become top-heavy at high SFRs or high star-forming densities \citep{guna11, weidner11} and increasing pre-GC cloud-core density \citep{marks12}. Furthermore, \citet{nara13} found that massive galaxies form the majority of their stars with a top-heavy IMF, but they may experience both top-heavy and bottom-heavy IMF phases during their life. The implications of a top-heavy IMF were studied extensively by \citet{dabringhausen09, dabringhausen10, dabringhausen12} and \citet{murray09}. Massive stars are then formed more frequently compared to a canonical IMF. They leave behind dark remnants such as neutron stars and black holes, which become visible in X-rays if they accrete matter from a low-mass companion star. \citet{dabringhausen12} found that these low-mass X-ray binaries are up to ten times more frequent in UCDs than expected for a canonical IMF.

\citet{weidner11} explored the implications of SFRs above $10^3~M_{\odot} \mathrm{yr}^{-1}$ on the mass function of SCs and found that either the ECMF becomes top-heavy or no low-mass SCs are formed. If the formation of low-mass SCs is indeed suppressed, our assumption of the lower mass limit, $M_{\mathrm{min}} = 5~M_{\odot}$, would not be justified. A change in $M_{\mathrm{min}}$ would have an effect on the total mass of each SC population, $M_{\mathrm{ECMF}}$ (Eq.~(\ref{M_ecmf})), and thus also on the SFR (Eq.~(\ref{sfr})). With a higher $M_{\mathrm{min}}$, $M_{\mathrm{ECMF}}$ will become lower so that the necessary SFR will decrease as well. However, it is difficult to quantify this effect since it is not clear what a more realistic lower mass limit would be.

\subsection{Results} \label{discuss_results}

In Sect.~\ref{sect_sample} we noted that the mass determination strongly depends on the modeled $M/L$ ratio: The mass estimate of GCs and UCDs by \citet{bc03} is on average between 15\% and 20\% lower than the one by \citet{maraston05} that we used. Based on the mass determination, this translates into an uncertainty in the SFR on the same order (cf.~Eq.~(\ref{sfr})), meaning that our derived SFRs are accurate to about 20\%.

Moreover, we extracted from our analysis for each run how many SC formation epochs (SCFE), $N_{\mathrm{SCFE}}$, are necessary to reproduce the observed GC/UCD sample, the total time this takes, $t_{\mathrm{SCFE,tot}}$, and the total stellar mass, $M_{\mathrm{tot}}$, formed during that time. All these values can be found in Table~\ref{tab_statistics} as a function of the parameter $t_4$ and the considered approach. The resulting values are similar, independent of the approach, and follow the same trend with increasing $\beta$ and $t_4$.

We assumed an age of 13~Gyr for our GC/UCD sample. Even though the age determination becomes more uncertain for older ages, we know from observations that the vast majority of objects in our sample formed more than 8~Gyr ago (cf.~Sect.~\ref{sect_sample}). Thus, the total SC formation time, $t_{\mathrm{SCFE,tot}} = N_{\mathrm{SCFE}} \cdot \delta t$ (Cols.~5, 8, 11) with $\delta t$, the length of one SC formation epoch, should not exceed several Gyr. This is the case for $\beta \lesssim 2.2$ mostly independent of $t_4$ and the chosen approach, and agrees nicely with the fact that $\beta \approx 2.0$ is typically found observationally \citep[e.g.,][]{zhang_fall99, lada_lada03, mccrady_graham07, chandar11}.

However, this does not imply that $\beta \gtrsim 2.2$ is ruled out. With increasing $\beta$, the corresponding values for $\delta t$ increase strongly up to about 80~Myr (cf.~Col.~3). Observationally, the duration of one SC formation epoch is found to be a few Myr up to a few tens Myr at most \citep[e.g.,][]{fukui99, yamaguchi01, tamburro08, egusa04, egusa09}. Apparently, for large $\beta$, our estimates for $\delta t$ are too high, which implies that the total SC formation time, $t_{\mathrm{SCFE,tot}}$, is an overestimate. Furthermore, calculating the total SC formation time through $t_{\mathrm{SCFE,tot}} = N_{\mathrm{SCFE}} \cdot \delta t$ assumes that the SC populations form consecutively. However, it is conceivable that SC formation could occur at the same time but at several separated places. In this case, an individual ECMF would be populated with SCs at each of those places, but the total SC formation time would be shorter than suggested by the formula. The latter might also explain why we find formation timescales several times longer than the estimate of shorter than 0.5~Gyr according to the downsizing picture \citep[e.g.,][see Sect.~\ref{discuss_formation}]{thomas99, recchi09}.

Today, the total stellar mass of NGC~1399 amounts to $6 \cdot 10^{11}~M_{\odot}$ within 80~kpc and rises to roughly $10^{12}~M_{\odot}$ at a distance of 670~kpc \citep[][their Table~1]{richtler08}. Recently, \citet{iodice16} derived a total stellar mass of NGC~1399 and its halo of about $6.6 \cdot 10^{11}~M_{\odot}$ , while the stellar mass in the halo amounts to about $4 \cdot 10^{11}~M_{\odot}$. These estimates set a strict upper limit on the total stellar mass, $M_{\mathrm{tot}}$, that formed during the entire SC formation process.

The $M_{\mathrm{tot}}$ columns of each approach in Table~\ref{tab_statistics} show that a disagreement with the above limit occurs again for $\beta \gtrsim 2.3$ mostly independent of $t_4$ and the considered approach. Two things should be mentioned here. First, the above limit regarding the total stellar mass does not imply that this mass is still stored in the GC/UCD sample today. Apparently, only a tiny portion still is: the present-day mass of our combined GC/UCD sample is about $5 \cdot 10^{9}~M_{\odot}$, which means about 1\% of the total stellar mass of NGC~1399,  but a large part of the initially formed SCs dissolved and contributed their stars to NGC~1399 and its halo. Moreover, even the surviving SCs lost part of their mass because of stellar evolution and the loss of stars in the tidal field, which also added to NGC~1399 and its surroundings. Second, the constraint on $M_{\mathrm{tot}}$ does not necessarily imply that larger $\beta$ have to be excluded: As mentioned in the previous subsection, the lower mass limit, $M_{\mathrm{min}}$, was assumed to be $5~M_{\odot}$ , which might be not well justified since very high SFRs as we derived here might prevent the formation of lower-mass SCs. If we indeed underestimate $M_{\mathrm{min}}$, then we overestimate the total mass of each SC population, $M_{\mathrm{ECMF}}$ (Eq.~(\ref{M_ecmf})), and therefore also the total mass ever formed in SCs, $M_{\mathrm{tot}}$. A higher $M_{\mathrm{min}}$ would lower all our estimates for $M_{\mathrm{tot}}$ in Table~\ref{tab_statistics}. However, the overestimation of the total mass would be highest for large $\beta$ since due to the steeper ECMF, more low-mass SCs are formed per high-mass SC.

As mentioned in Sect.~\ref{sect_decomposition}, we introduced a margin of five percent when replicating the observed GC/UCD mass distribution. For comparison, we checked how no tolerance at all and a margin of twenty percent influences the outcome: the higher the margin, the more the generated GC/UCD mass distributions is shifted downward at the low-mass end. Consequently, when the mass function contains fewer SCs, slightly fewer formation epochs and lower SFRs are required. However, since mainly the low-mass end of the GC/UCD distribution is affected but not the high-mass end, the high-SFR end of the SFR distribution does not change. The most noticable difference appears in the total mass, $M_{\mathrm{tot}}$, of all SCs ever formed: For a larger margin, slightly fewer formation epochs are necessary. Large $\beta$ are affected the most because for them, the relative number of low-mass SC, produced during every formation epoch, is higher. As compared to a margin of five percent, the difference in $M_{\mathrm{tot}}$ varies between $+$2 \% and $-$8 \% for $\beta = 1.7$ and $+$5 \% and $-$15 \% for $\beta = 2.5$ for no tolerance at all and a margin of twenty percent, respectively. To avoid underestimating the total mass, $M_{\mathrm{tot}}$, we chose a relative small margin of 5 \%.

In summary, it appears that all cases up to $\beta \approx 2.2$ are in agreement with the conditions set by NGC~1399 mostly independent of $t_4$, the considered approach, and the choice of the margin. Again, this fits the observations of young SCs well, where usually $\beta \approx 2.0$ is found. However, as already mentioned in Sect.~\ref{sect_sfr.distr}, we do not regard the standard approach as very well justified because according to it, all GCs/UCDs are assumed to be genuine SCs, even though in numbers, the derived values for $t_{\mathrm{SCFE,tot}}$ and $M_{\mathrm{tot}}$ do not vary much among the different approaches (cf.~Table~\ref{tab_statistics}).

\subsection{Formation of NGC~1399 and its GC/UCD system} \label{discuss_formation}

Here, we discuss possible formation scenarios of elliptical galaxies that can explain SFRs of more than 1000~$M_{\odot} \mathrm{yr}^{-1}$. There are suggestions that a massive elliptical might be the result of a merger of two (gas-rich spiral) galaxies \citep[e.g.,][]{lilly99, eser14} and subsequent accretion of additional galaxies in the course of time. This is based on the observations of so-called (ultra-/hyper-)luminous infrared galaxies ((U/H)LIRGs), which are characterized by a substantial emission in the infrared (LIRGs: $L_{\mathrm{IR}} > 10^{11} L_{\odot}$; ULIRGs: $L_{\mathrm{IR}} > 10^{12} L_{\odot}$; HLIRGs: $L_{\mathrm{IR}} > 10^{13} L_{\odot}$). The energy behind these high luminosities is produced by active galactic nuclei (AGN) and/or an intense starburst \citep[e.g.,][]{carico90, condon91} and then re-radiated in infrared wavelengths. Observations show that the relative contribution of the AGN and the starburst component to the IR emission is a function of the luminosity: LIRGs and low-luminosity ULIRGs are mostly powered by a starburst, while high-luminosity ULIRGs and HLIRGs are dominated by an AGN \citep[e.g.,][]{veilleux99, nardini10}.

In particular for ULIRGs and partly for HLIRGs, the picture emerged that they depict the merger of two gas-rich galaxies \citep[e.g.,][]{genzel00, farrah02}, thereby producing an elliptical galaxy \citep[e.g.,][]{kormendy92, genzel01}. A strong encounter instead of a merger between two gas-rich galaxies may be what is observed, however (see \citealt{kroupa15review} for a discussion on mergers vs. interactions), and in this case, the ULIRGs/HLIRGs may not be progenitors of elliptical galaxies. Whether a merger or not, the interaction triggers an intense star formation with SFRs between roughly 200 and 4000 $M_{\odot} \mathrm{yr}^{-1}$ \citep[e.g.,][]{farrah02, lefloch05, takata06, bastian08, ruiz13, eser14}, which matches the range of SFRs we find well. These galaxies cover a wide redshift range of up to $z \approx 3$ (i.e.,~light emission up to 11.5~Gyr ago) and show a tendency of higher SFRs with higher redshifts (e.g., cf.~\citealt{bastian08, eser14} vs.~\citealt{farrah02, takata06}, see also \citealt{rr00}), which is confirmed by studies focusing on the evolution of the SFR with redshift or over cosmic time \citep[e.g.,][]{lefloch05, schiminovich05, speagle14, mancuso16}.

Toward higher redshifts, the level of obscuration of galaxies increases: They are usually barely visible in the optical and UV, but have an enormous emission at far-IR and sub-mm wavelengths \citep[e.g.,][]{michalowski10, nara15, mancuso16}. This is why they are called sub-millimeter galaxies (SMGs) or, more generally, dusty, star-forming galaxies (DSFGs). According to their IR-luminosities, the most luminous of them fall into the regime of HLIRGs and are among the most luminous, heavily star-forming galaxies in the Universe \citep[e.g.,][]{michalowski10, hainline11, casey14}. Their SFRs are similar to those of ULIRGs/HLIRGs and typically lie between a few hundred and a few thousand $M_{\odot} \mathrm{yr}^{-1}$ \citep[e.g.,][]{zadeh12, swinbank14, cunha15, simpson15}, with the tendency that SMGs with higher redshifts have higher SFRs \citep{cunha15}. It has been argued that these DSFGs are not major mergers but galaxies experiencing their highest star formation activities \citep[e.g.,][]{farrah02, nara15, mancuso16}. This matches with the fact that massive elliptical galaxies are known to be $\alpha$-element enhanced and metal-rich, such that they cannot have formed from the mergers of pre-existing comparatively metal-poor disk galaxies.

Moreover, these galaxies are extremely massive: they can be interpreted to reside in particle dark matter or in phantom dark matter halos \citep[e.g.,][]{famaey12, lueghausen15} with masses of between more than $10^{11}$ and more than $10^{13}~M_{\odot}$ \citep{hickox12, bethermin13} and have stellar masses in the range of between lower than $10^{11}$ and more than $10^{12}~M_{\odot}$ \citep[e.g.,][]{swinbank06, michalowski10, hainline11}. For halo masses around $10^{12} M_{\odot}$, star formation occurs most efficiently \citep[e.g.,][]{behroozi13, bethermin13, wang13}, which is why the progenitors of even more massive present-day halos passed through this mass quickly, and thus formed most of their stars on timescales shorter than 1-2 Gyr \citep[see left panel of their Fig.~13 in][]{behroozi13, marsan15}. 

Overall, DSFGs represent a phase in massive galaxy evolution that marks the transition from cold gas-rich, heavily star-forming galaxies to passively evolving systems \citep{hickox12}. In simulations by \citet{nara15}, this active phase was accompanied by a significant build-up of stellar mass; thereafter, these galaxies are expected to evolve into massive ellipticals \citep[e.g.,][]{michalowski10, hickox12}. For instance, when passively evolving high-redshift SMGs to the present time, \citet{hainline11} found their luminosity (and therefore mass) distribution to be similar to that of massive ellipticals in the Coma galaxy cluster. The authors noted that typical SMGs cannot represent the formation phase of the very luminous cD-type galaxies observed in galaxy clusters because the baryonic mass of a typical SMG is too low. \citet{miller15} emphasized that there are better tracers for the assembly of the most massive structures in the Universe than SMGs, Lyman-break galaxy analogs, for instance.

However, it needs to be taken into account that the Coma galaxy cluster is much richer in galaxies and in mass than the Fornax galaxy cluster. The center of the Coma galaxy cluster is dominated by two giant ellipticals, NGC~4874 and NGC~4889. Measurements by \citet[with $h=0.678$ from \citealt{planck14}]{okabe10} and \citet{andrade13} resulted in halo masses of $6.7 \cdot 10^{12}~M_{\odot}$ and $7.6 \cdot 10^{12}~M_{\odot}$ for NGC~4874, and $11.4 \cdot 10^{12}~M_{\odot}$ and $9.1 \cdot 10^{12}~M_{\odot}$ for NGC~4889, respectively. In comparison, \citet[their Table~1]{richtler08} estimated a halo mass of $3.5 \cdot 10^{12}~M_{\odot}$ for NGC~1399 within a radius of 80~kpc. Clearly, NGC~1399 is several times less massive than the center of the Coma galaxy cluster and could be considered as a less massive version of it at most. Nevertheless, there are hints that NGC~1399 went through multiple interactions that left an imprint on the spatial distribution of GCs \citep{dabrusco16} and a faint stellar bridge in the intracluster region on the west side of NGC~1399 \citep{iodice16}.

Regarding the formation timescale, massive elliptical galaxies like NGC~1399 must have formed on a short timescale, while less-massive galaxies are known to have formed over longer times, which is known as downsizing \citep[e.g.,][]{cowie96, thomas99, juneau05, recchi09}. Assuming that most of the mass of NGC~1399 was formed early on, the build-up must have been completed within less than 0.5~Gyr according to Eq.~(19) in \citet[][see also their Fig.~18]{recchi09}. This allows us to estimate the SFR during the formation: to simplify matters, we can assume a constant SFR over that time, which leads to $\mathrm{SFR} = M / t \approx 5 \cdot 10^{11}~M_{\odot} / 0.5~\mathrm{Gyr} \approx 1000~M_{\odot}/\mathrm{yr}$. Since the formation timescale might be shorter and the SFR over the formation period does not have to be constant, the peak SFRs might be a few times higher than estimated in this simple calculation.

In summary, the range of SFRs we found in Sect.~\ref{sect_sfr.distr} is in very good agreement with the SFRs observed in SMGs and the simple estimate based on downsizing. Moreover, the age of the GCs/UCDs (Sect.~\ref{sect_sample}) sets a limit on the formation timescale of NGC~1399 since the central Fornax galaxy and its GC/UCD system formed probably coevally. Most of the mass was built up within a few Gyr, which matches the generally short formation timescales of massive elliptical galaxies. In that respect, it seems reasonable to assume that as a result of their extreme star formation activity, massive high-redshift SMGs might represent the progenitors of cD galaxies like NGC~1399.

\section{Conclusions} \label{sect_concl}

We combined a spectroscopic and a photometric sample of GCs/UCDs around the giant elliptical NGC~1399 in the Fornax galaxy cluster to derive their overall mass distribution (Sect.~\ref{sect_sample}) and corrected it for stellar and dynamical evolution (Sect.~\ref{sect_corrections}) to obtain their mass function at the time of formation. Then, this "natal" mass function was decomposed into individual SC populations distributed according to the ECMF (Sect.~\ref{sect_decomposition}, schematic plot in Fig.~\ref{fig_decompo_sketch}). The upper mass limit of each population, $M_{\mathrm{max}}$, was converted into an SFR according to the SFR-$M_{\mathrm{max}}$ relation \citep{weidner04} as was fit in \citet{schulz15} (Sect.~\ref{sect_sfr.distr}). The resulting SFR distributions (Fig.~\ref{fig_sfr}) reveal which SFRs are required to form the entire GCs/UCDs system around NGC~1399. 

When restoring the natal GC/UCD mass function, we assumed different lifetimes of the SCs, parameterized by $t_4$, to account for the variable strength of the tidal field depending on the distance to the center of NGC~1399 (Sect.~\ref{sect_corrections}). Moreover, we used three different approaches regarding the treatment of the natal GC/UCD mass function. First, in the standard approach we assumed that all GCs/UCDs were formed in an SC formation process and are therefore ancient SCs. Second, since at least the most massive UCD in our sample, UCD3, shows hints of being a merged SC or the nucleus of a stripped dwarf galaxy, we excluded this object from our sample. Third, \citet{pfeffer14} derived a possible distribution of dwarf galaxies whose envelopes were stripped away. We assumed that all objects in our GC/UCD mass function compatible with their distribution are stripped nuclei so that only the remaining GCs/UCDs need to be replicated by our method. All these modified samples were then treated as described above, meaning that they were decomposed into individual SC populations from which a distribution of SFRs was deduced based on $M_{\mathrm{max}}$ of each population.

Although we made different assumptions regarding the lifetime of the GCs/UCDs and modified the original GC/UCD sample, the outcome was mainly determined by the index $\beta$ of the underlying ECMF, while the influence of the parameter $t_4$ or the modified sample was rather of second order. We extracted from our analysis for each combination of parameters the distribution of SFRs required to build up the observed GC/UCD sample (Fig.~\ref{fig_sfr}), the time this takes, $t_{\mathrm{SCFE,tot}}$, and the total stellar mass, $M_{\mathrm{tot}}$, formed during that time (Table~\ref{tab_statistics}). We found that

\begin{itemize}
\item the results for $t_{\mathrm{SCFE,tot}}$ and $M_{\mathrm{tot}}$ are well within the constraints set by the age of the GCs/UCDs and the total stellar mass of NGC~1399 for $\beta \lesssim 2.2$,
\item the favored values for $\beta$ nicely fit the observations of young SCs where usually $\beta \approx 2.0$ is measured, and
\item the peak SFRs derived by us agree well with the range of SFRs observed in massive high-redshift SMGs and also with an estimate based on downsizing.
\end{itemize}

As discussed in Sect.~\ref{sect_discussion}, our assumption for the lower mass limit of SCs, $M_{\mathrm{min}} = 5~M_{\odot}$, might be an underestimate, given that high SFRs as derived here might suppress the formation of very low-mass SCs. Moreover, the length of one SC formation epoch, $\delta t$, that we used for large $\beta$ is higher than observed in star-forming regions. Increasing $M_{\mathrm{min}}$ and, for large $\beta,$ decreasing $\delta t$ would lead to a shorter total time for SC formation, $t_{\mathrm{SCFE,tot}}$, and to a lower total stellar mass, $M_{\mathrm{tot}}$, so that even higher $\beta$ could still be in agreement with the observational constraints in NGC~1399. Additionally, we regard the standard approach, where all GCs/UCDs are assumed to be genuine SCs, as not very well justified because UCD3, the most massive object in our sample, is clearly not a typical GC. However, except for the higher peak SFRs, caused by UCD3, the outcome from the standard approach does not differ much from the other two approaches.

In conclusion, NGC~1399 might have originated from an intense starburst similar to those observed in massive SMGs in the distant Universe. During that starburst, in particular the most massive GCs/UCDs were formed along with many lower mass GCs within a few Gyr. The dissolution and tidal disruption of a part of the GCs/UCDs probably fed the build-up of NGC~1399 and its halo, while a part of the GCs/UCDs was able to survive, allowing us to observe them today.

However, here, the GC/UCD sample was analyzed in its entirety without differentiating the red and blue subsamples, that is, the metal-rich and metal-poor GCs/UCDs. In a future paper we will reapply the method described here to the two subsamples to see whether and by how much the formation conditions of the red and blue GCs/UCDs differ.


\bibliographystyle{aa} 
\bibliography{/home/cschulz/Documents/essentials/d3-aa}   

\end{document}